\documentclass[aps,prb,twocolumn,superscriptaddress]{revtex4-1}
\usepackage{hyperref}
\usepackage{dsfont}
\usepackage{amsmath}
\usepackage{amssymb}
\usepackage{graphicx}
\usepackage{color}

\newcommand{\pp}[1]{p_{#1}^{\prime}}      
\newcommand{\ppb}[1]{\mathbf{p}_{#1}^{\prime}} 

\newcommand{\pb}[1]{\mathbf{p}_{#1}}  
\newcommand{\Pb}{\mathbf{P}}  

\newcommand{\nb}{\mathbf{n}} 
\newcommand{\nbp}[1]{\mathbf{n}_{#1^{\prime}}} 

\newcommand{\np}[1]{n_{#1^{\prime}}} 

\newcommand{\ef}{\epsilon_{\mathrm F}} 
\newcommand{\pf}{p_{\mathrm F}} 
\newcommand{\pft}{\tilde{p}_{\mathrm{F}}} 
\newcommand{\nf}{n_{\mathrm{F}}} 
\newcommand{\Deff}{\Delta_{\mathrm{eff}}} 

\newcommand{\diff}{d}

\begin{document}


\title{Long lifetimes of ultrahot particles in interacting Fermi systems}


\author{M. Bard}
\affiliation{Institut f\"ur Nanotechnologie, Karlsruhe Institute of Technology, 76021 Karlsruhe, Germany}
\author{I.~V. Protopopov}
\affiliation{Department of Theoretical Physics, University of Geneva, 1211 Geneva, Switzerland  }
\affiliation{Landau Institute for Theoretical Physics, 119334 Moscow, Russia}
\author{A.~D. Mirlin}
\affiliation{Institut f\"ur Nanotechnologie, Karlsruhe Institute of Technology, 76021 Karlsruhe, Germany}
\affiliation{Institut f\"ur Theorie der Kondensierten Materie, Karlsruhe Institute of Technology, 76128 Karlsruhe, Germany} 
\affiliation{Petersburg Nuclear Physics Institute, 188350 St. Petersburg, Russia}


\date{\today}

\begin{abstract}

The energy dependence of the relaxation rate of hot electrons due to interaction with the Fermi sea is studied. 
We consider 2D and 3D systems, quasi-1D quantum wires with multiple transverse bands, as well as single-channel 1D wires. Our analysis includes both spinful and spin-polarized setups, with short-range and Coulomb interactions. We show that, quite generally, the relaxation rate is a non-monotonic function of the electron energy and decays  as a power-law at high energies. In other words, ultra-hot electrons regain their coherence with increasing energy.  Such a behavior was observed in a recent experiment on multi-band quantum wires, J. Reiner et al, Phys. Rev. X {\bf 7}, 021016 (2017).

\end{abstract}


\maketitle


\section{Introduction}

Relaxation of excitations due to interaction with the surrounding system is a fundamental phenomenon that governs the behavior of complex many-body  systems at sufficiently large time scales and is in the core of thermalization and ergodization.  Further, relaxation breaks the quantum coherence and curtains quantum laws of nature, thus giving rise to the largely classical world that surrounds us. 

Already in the early days of the quantum many-body theory it was realized, however, that the relaxation times associated with the {\it low-energy} excitations in the many-body systems can be long (compared to other microscopic time scales). 
The paradigmatic example of such a situation is the Landau Fermi-liquid theory  that rests on the fact that, in an interacting electronic system, a Landau quasiparticle with excitation energy $\epsilon$ (counted from the Fermi energy $\ef$)  has the decay rate (inverse lifetime) $1/\tau_\epsilon\propto \epsilon^2/\ef\ll \epsilon$.   The long lifetimes of quasiparticles at low energies give rise to a plethora of quantum phenomena in the low-temperature properties of electronic systems observed in a variety of experiments, including, e.g. quantum corrections to conductivity \cite{WeakLocalization} and quantum Hall interferometry \cite{QHInterferometry}.

In view of the physical importance of relaxation processes, they were extensively studied over decades in various 
condensed matter systems and for various types of excitations, including electrons in normal metals\cite{EERelaxationNormal},
Bogolyubov quasiparticles in superconductors\cite{EERelaxationSuperconductors} and Bose gases\cite{RelaxationColdAtoms,Beliaev,Tan2010,Ristivojevic2014,Ristivojevic2016}, electrons in one-dimensional (1D) quantum wires\cite{Relaxation1DExp} and quantum Hall edge channels\cite{RelaxationHallExp}, to name a few. Most of these studies focused on the low-energy excitations that are usually {\it expected} to relax slowly and whose properties in many cases show remarkable degree of universality. 

The fact that not only at low energies can fermionic excitations exhibit long relaxation times  was emphasized recently in 
Ref. \onlinecite{Beidenkopf2017}. This work studied the interaction-induced decoherence of hot electrons in a semiconducting nanowire by means of scanning tunneling microscopy. Remarkably, it was observed that, while at low energies the electron relaxation time reduces as its energy grows (the behavior familiar from the conventional Fermi-liquid theory), it becomes long again  at excitation energies larger than the Fermi energy. In other words, 
electrons regain their coherence at high energies.  This surprising behavior was explained in Ref.~\onlinecite{Beidenkopf2017}  by an analysis of the relaxation in processes that involve only electrons from the lowest band of the transverse quantization in the nanowire. 

The results of Ref.~\onlinecite{Beidenkopf2017} pose natural questions. How universal is the non-monotonic behavior of the relaxation rate and the regain of coherence at high energies? In particular, do they persist in quasi-1D systems with several transverse bands involved in relaxation? 
Are they also relevant to two-dimensional (2D) and three-dimensional (3D) systems? 

The purpose of this work is to answer these questions.  We show that the non-monotonicity of the relaxation rate and the regain of quantum coherence is a very general and robust phenomenon.  We do this by studying several models of interacting fermions, ranging from fermions with isotropic spectrum in $D\geq 2$,  through multi-band quantum wires, to strictly 1D interacting fermions. 
We show that in all these models the electron relaxation rate decays as a power law with the momentum $p_1$ of the hot fermion, provided that the interparticle interaction decays sufficiently fast as a function of the momentum transfer. In 2D and 3D as well as in quasi-1D the relaxation rate scales as $1/p_1$ while in 1D the rate vanishes as $1/p_1^5$.  In this analysis, we focus on models with the interaction potential in momentum domain, $V(q)$, characterized  by a single momentum scale $q_0$ such that the interaction can be expanded, $V(q)=V_0(1-q^2/q_0^2)$, for $q\ll q_0$ and gets suppressed sufficiently strongly at $q\gg q_0$. We also explore the case of Coulomb interaction and demonstrate that the non-monotonic behavior of the relaxation rate applies in this situation as well.

A related problem of decay of high-energy \emph{bosonic} quasiparticles was recently studied in Refs. \onlinecite{Tan2010,Ristivojevic2016} in 1D geometry. The authors of these works found a saturation of the decay rate at high energy, which is in contrast to the $1/p_1^5$-decay that we predict for fermionic quasiparticles. We will return to the origin of this difference in Sec.~\ref{Sec:Summary}.

The structure of the paper is as follows. We start  in Sec. \ref{Sec:2D-3D} with the simplest possible setting: weakly interacting fermions with an isotropic parabolic dispersion in $D\geq 2$ spatial dimensions.  We analyze two-particle collision processes and show that 
 the corresponding relaxation rate is a non-monotonic function of the quasiparticle energy.  
 section \ref{Sec:MultiBand} is devoted to the discussion of the relaxation in multi-band 
 metallic wires: a situation that can be viewed as a strongly anisotropic limit of 2D and 3D models of 
 Sec. \ref{Sec:2D-3D}. 
  In Sec. \ref{Sec:1D} we consider fermions in a 1D parabolic band 
 where two-particle collisions are forbidden by energy and momentum conservation laws and the analysis of three-fermion collisions is required.  Our results are summarized in Sec. \ref{Sec:Summary}.



\section{Isotropic 3D and 2D cases}
\label{Sec:2D-3D}

The simplest setting one can assume to study the relaxation in a condensed matter system at high energies is that of (spinless or spinful) particles with parabolic dispersion in $(D\geq 2)$-dimensional space. Clearly, the cases $D=2$ and $D=3$ are of particular interest from the physical point of view. We thus consider in this section 
an isotropic $D$-dimensional Fermi sea with Fermi momentum $\pf$ and, on top of it, an electron with momentum $p_1 \gg \pf$. We assume that the particles interact via an interaction $V(q)$ characterized by a single momentum scale $q_0$, so that $V(q)$ can be expanded,
$V(q)\sim V_0(1-q^2/q_0^2)$,  for $q\lesssim q_0$ and is sufficiently strongly suppressed at  $q\gg q_0$.  As a guiding example, one can think about a model interaction with an exponential decay, e.g.,  $V(q) = V_0 e^{-(q/q_0)^2}$. As we discuss in the end of this section,  our results remain applicable also in the case of screened Coulomb interaction in 2D and 3D.  We assume $q_0\gtrsim \pf$, while the relation between $q_0$ and $p_1$ can be  arbitrary.

We are interested in the relaxation rate for our hot particle which is given by the Fermi golden rule,
\begin{widetext}
\begin{equation}
\frac{1}{\tau_{p_1}}=\frac{1}{2!}\int  d\pb{2} d\ppb{1}d \ppb{2}\,\delta\left(E_{\rm i}-E_{\rm f}\right)
\delta\left({\bf P}_{\rm i}-{\bf P}_{\rm f}\right) \nf(\epsilon_2)[1-\nf(\epsilon_{1}^{\prime})][1-\nf(\epsilon_{2}^{\prime})]
\left|M_{\pb{1},\pb{2}}^{\ppb{1},\ppb{2}}\right|^2.
\label{Eq:relaxation-rate-2D-3D}
\end{equation}
\end{widetext}
This expression contains only the out-scattering rate since by assumption the initial distribution consists of a filled Fermi sea and an additional high-energy fermion.
Here  $\pb{1}$ and $\pb{2}$ ($\ppb{1}$ and $\ppb{2}$) are the particle momenta before (after) the scattering, 
$E_{\rm i}$, ${\bf P}_{\rm i}$ and $E_{\rm f}$, ${\bf P}_{\rm f}$ are the total energy and momentum of the two particles before and after the collision, $\delta$ functions express the energy and momentum conservation,
 and we use the notation $\epsilon_i\equiv\epsilon_{\pb{i}}=\pb{i}^2/2m$. 
 The matrix element 
$M_{\pb{1},\pb{2}}^{\ppb{1},\ppb{2}}$ consists of the direct and exchange terms. In the spin-polarized case we have
\begin{equation}
\left|M_{\pb{1},\pb{2}}^{\ppb{1},\ppb{2}}\right|^2=\left[V(\left|\pb{1}-\ppb{1}\right|)-V(\left|\pb{1}-\ppb{2}\right|)\right]^2.
\end{equation}
In the case of fermions with spin we find, after the summation over the spin polarization of the second electron:
\begin{equation}
\begin{split}
\left|M_{\pb{1},\pb{2}}^{\ppb{1},\ppb{2}}\right|^2=
\left[V(\left|\pb{1}-\ppb{1}\right|)\right]^2+\left[V(\left|\pb{1}-\ppb{2}\right|)\right]^2\\
-V(\left|\pb{1}-\ppb{1}\right|)V(\left|\pb{1}-\ppb{2}\right).
\end{split}
\end{equation}
 
Exploiting rotational invariance of the problem, one can recast Eq. (\ref{Eq:relaxation-rate-2D-3D}) into a 
convenient form (see Appendix~\ref{App:2D_and_3D}):
\begin{widetext}
\begin{equation}
\begin{split}
\frac{1}{\tau_{p_1}}=\frac{ S_{D-2} }{2p_1^{D-2}}\int_{0}^{\infty} dP P^{D-1}\int_{\left|P/2-p_1\right|}^{P/2+p_1}q^{2D-3} dq 
\int_0^\pi d\phi\, d\phi^\prime\, \left(\sin\phi \sin\phi^\prime\right)^{D-2}&\delta\left(\epsilon_1 -\frac{p_1^2}{2m}\right)  
\\\times &\nf(\epsilon_2)[1-\nf(\epsilon_{1}^{\prime})][1-\nf(\epsilon_{2}^{\prime})]w_q(\phi, \phi^\prime).
\end{split}
\label{Eq:relaxation-rate-2D-3D-convenient}
\end{equation}
 \end{widetext}
 Here $S_d$ is the volume of a $d$-dimensional sphere ($S_{d=0}=2$). 
Physically, the integration variables  $P$ and $q$ in Eq. (\ref{Eq:relaxation-rate-2D-3D-convenient}) are the 
center-of-mass and relative momentum of the particles in the collision process, $P=\left|{\bf p}_1+{\bf p}_2\right|=
\left|{\bf p}_1^\prime +{\bf p}_2^\prime\right|$ and $2q=\left|{\bf p}_1-{\bf p}_2\right|=
\left|{\bf p}_1^\prime -{\bf p}_2^\prime\right|$; angle $\phi$ ($\phi^\prime$) is the angle
between center-of-mass momentum and relative momentum before (after) the collision. The energies $\epsilon_i$ and $\epsilon_i^\prime$ are functions of integration variables given by
\begin{eqnarray}
2m\epsilon_{1, 2}= \frac{P^2}{4}+q^2\pm Pq \cos\phi,\\
2m\epsilon^\prime_{1,2}= \frac{P^2}{4}+q^2\pm Pq \cos\phi^\prime.
\end{eqnarray} 
The $\delta$ function in Eq. (\ref{Eq:relaxation-rate-2D-3D-convenient}) ensures that the momentum of one of the incoming particles equals $p_1$ and fixes the angle $\phi$ to 
\begin{equation}
\phi_0=\arccos \frac{p_1^2-q^2-P^2/4}{P q}.
\label{Eq:phi0}
\end{equation}
The limits of the $q$-integration guarantee that $0<\phi_0<\pi$. Finally, the function $w_q(\phi, \phi^\prime)$ represents  the (properly angle-averaged) matrix element squared. It is given by 
\begin{equation}
w_q(\phi, \phi^\prime)=S_{D-3}\int_0^\pi d\gamma  \left(\sin\gamma\right)^{D-3} \left[V_+-V_-\right]^2
\label{Eq:wDef}
\end{equation}
and
\begin{equation}
w_q(\phi, \phi^\prime)=S_{D-3}\int_0^\pi d\gamma  \left(\sin\gamma\right)^{D-3} \left[V_+^2+V_-^2-V_+V_-\right]
\label{Eq:wDefSpin}
\end{equation}
in the spinless and spinful cases, respectively, with
\begin{equation}
V_{\pm}=V\left[ q
\sqrt{2\left(1\pm\cos\phi\cos\phi^\prime\pm\cos\gamma \sin\phi\sin\phi^\prime\right)}\right].
\label{Eq:Vpm}
\end{equation}
For the special case of $D=2$, the integration over $\gamma$ in Eqs. (\ref{Eq:wDef}) and (\ref{Eq:wDefSpin}) should be understood according to 
\begin{equation}
S_{-1}\int \frac{d\gamma}{\sin\gamma}\longrightarrow\sum_{\gamma=0, \pi}.
\end{equation}

\begin{figure}
\includegraphics[width=220pt]{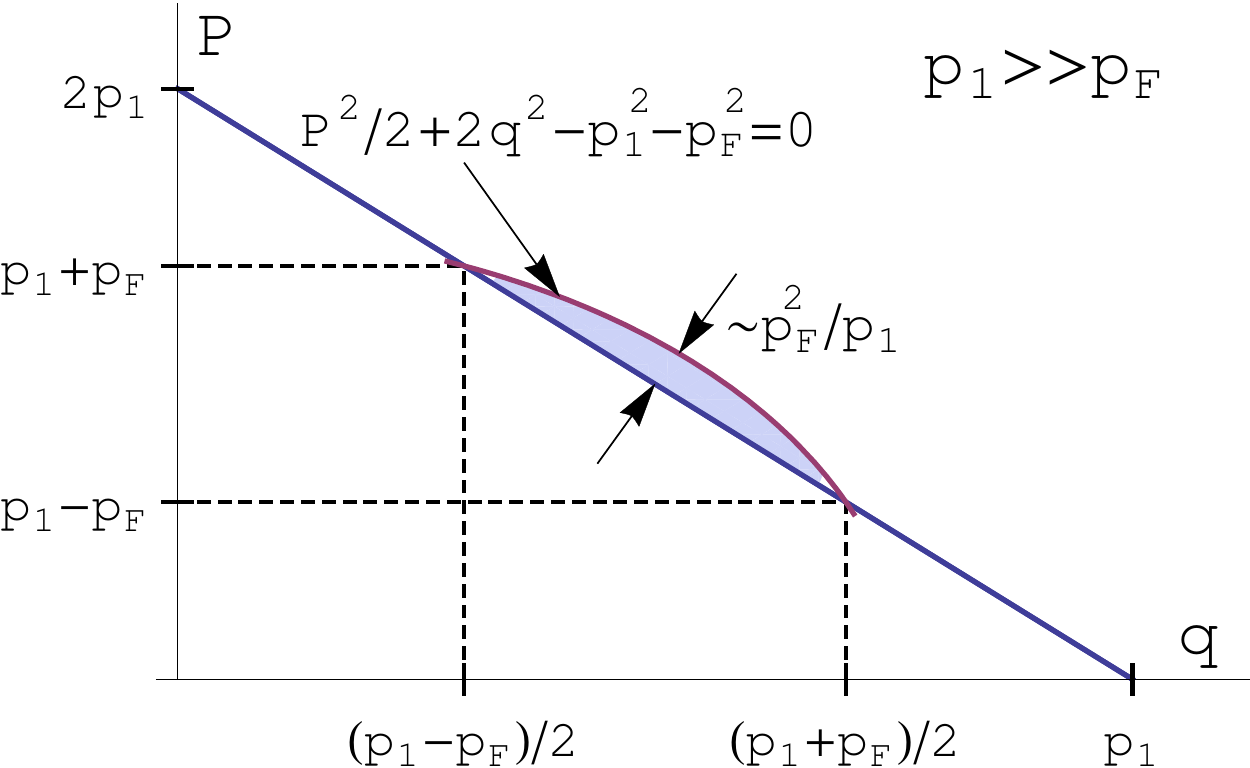}
\caption{High-energy scattering process ($p_1 \gg \pf$)  in terms of the center-of-mass and relative momentum of the colliding particles. The Fermi distribution $\nf(p_2)$ restricts the allowed values of $q$ and $P$ to a small vicinity of the point $(p_1/2, p_1)$, which is represented by a shaded region  between the line $P/2+q=p_1$ and the ellipse $P^2/2+2q^2-p_1^2-\pf^2=0$.  }
\label{Fig:PQPlane1}
\end{figure}

Equation (\ref{Eq:relaxation-rate-2D-3D-convenient}) is fully general and applies to arbitrary temperature and momentum $p_1$. It simplifies considerably in the case of $T=0$ and $p_1\gg \pf$. Under these conditions, the Fermi function $\nf(\epsilon_2)$ restricts the integration over $p$ and $q$ to a small vicinity of the point $P=p_1$, $q=p_1/2$, see Fig. \ref{Fig:PQPlane1}.  
The angle $\phi_0$ defined by Eq. (\ref{Eq:phi0}) is then small in the whole range of integration,
\begin{equation}
 \phi_0 \simeq 2\frac{\sqrt{P+2q-2p_1}}{\sqrt{p_1}}\lesssim \frac{\pf}{p_1} \ll 1. 
\end{equation}
In contrast to the case of low-energy scattering processes familiar from the Fermi-liquid theory, the Fermi factors associated to the outgoing momenta, $(1-\nf(\epsilon_i^\prime))$ do not play a major role here, as the typical momenta after the scattering are large. 
Under the above assumptions, the effect of these factors is only  to exclude almost perfect forward scattering processes characterized by 
\begin{equation}
\min(\phi^\prime, \pi-\phi^\prime)< \phi_0^\prime=\arccos\frac{p^2/4+q^2-\pf^2}{p q}\lesssim \frac{\pf}{p_1}.
\end{equation}

Under the assumption that the characteristic momentum transfer $q_0$ is larger than (or of the order of) the Fermi momentum, $q_0\gtrsim \pf$,   the smallness of the angles $\phi_0$ and $\phi_0^\prime$ allows us to fully decouple the $\phi^\prime$ integration in Eq. (\ref{Eq:relaxation-rate-2D-3D-convenient}). This yields
\begin{equation}
\frac{1}{\tau_{p_1}}\sim m p_1^{D-2}\pf^D\int_0^\pi d\phi^\prime \left(\sin\phi^\prime\right)^{D-2}w_{p_1/2}(0, \phi^\prime).
\label{Eq:relaxation-rate-2D-3D-semifinal}
\end{equation}
Here we have taken into account the characteristic value of the angle $\phi_0\sim \pf/p_1$ as well as the available area in the $(q, P)$-plane, $\pf^3/p_1$.

In Eq.~(\ref{Eq:relaxation-rate-2D-3D-semifinal}) and in analogous formulas below, the symbol ``$\sim$'' means ``equal up to a numerical coefficient of order unity''. This coefficient depends on the specific model of the interaction. 

The resulting scaling of the relaxation time (\ref{Eq:relaxation-rate-2D-3D-semifinal}) depends on the presence of spin as well as on the relation between $p_1$ and $q_0$. 
At $\pf\ll p_1\ll q_0$, the function $w_{p_1/2}$  in Eq. (\ref{Eq:relaxation-rate-2D-3D-semifinal}) can be expanded in powers of $p_1^{-1}$. For fermions with spin, the function $w_{p_1/2}$  is given by Eq.~(\ref{Eq:wDefSpin}), which
 leads to the estimate $w_{p_1/2}\sim V_0^2$ and results in
the collision rate 
\begin{equation}
\frac{1}{\tau_{p_1}}\sim m V_0^2 \pf^D p_1^{D-2}\,,\qquad   \pf\ll p_1\ll q_0.
\label{Eq:rate-isotropic-large-q0-spin}
\end{equation}
The result  (\ref{Eq:rate-isotropic-large-q0-spin}) is determined merely by the phase space available for the collision process. 
On the other hand, for spinless (or spin-polarized) fermions in the same parameter regime, the leading contribution to $w_{p_1/2}$ vanishes due to Hartree-Fock cancellation, see Eq. (\ref{Eq:wDef}). This results in the suppression of the relaxation rate 
(compared to the spinful case), yielding
\begin{equation}
\frac{1}{\tau_{p_1}}\sim m V_0^2 \frac{\pf^D p_1^{D+2}}{q_0^4}\,,\qquad   \pf\ll p_1\ll q_0.
\label{Eq:rate-isotropic-large-q0-spinless}
\end{equation}

\begin{figure}
\includegraphics[width=0.95\columnwidth]{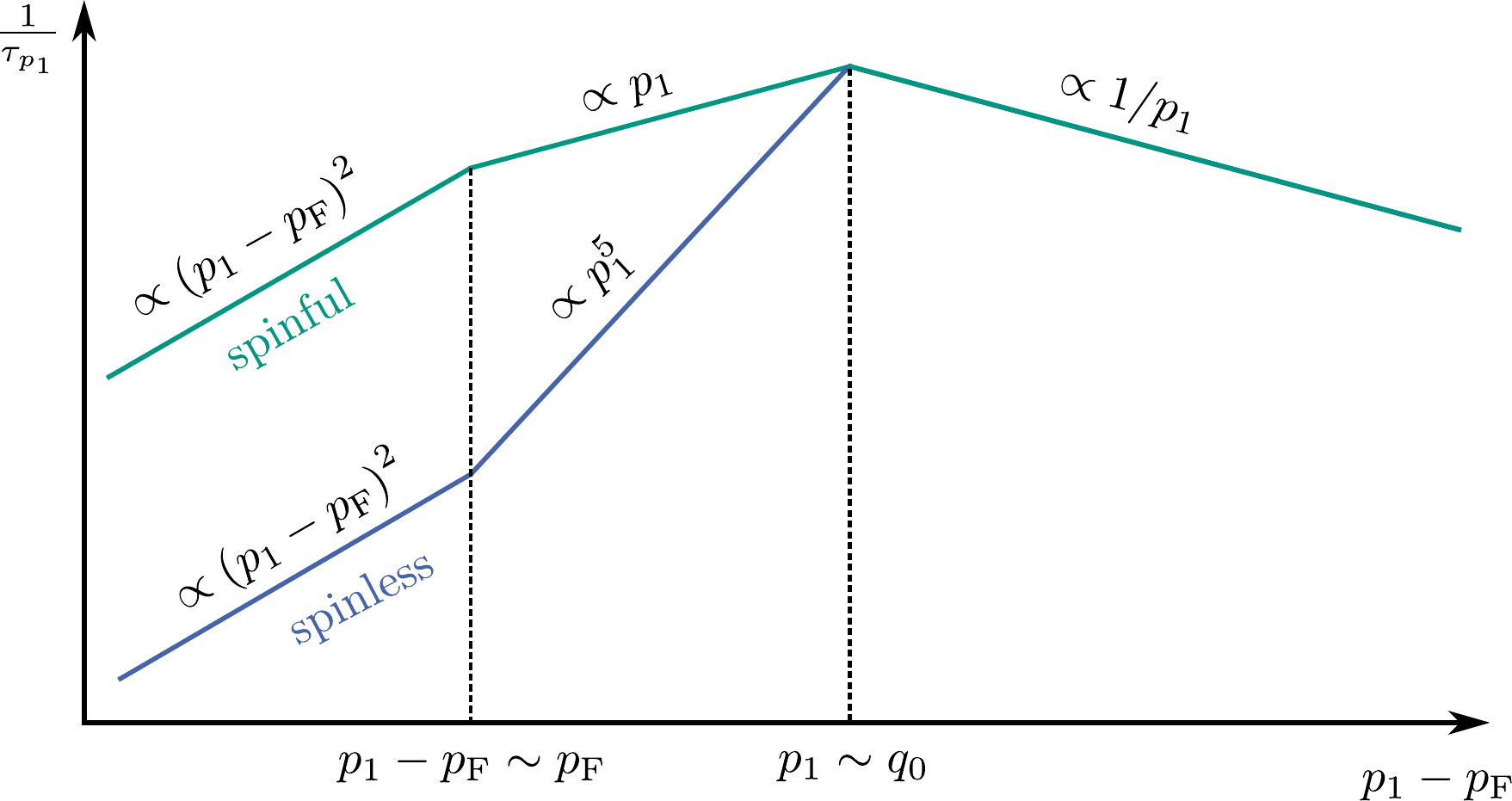}
\caption{Schematic plot (on log-log scale) of the  momentum dependence of relaxation rate in $D=3$, as predicted by 
Eqs.  (\ref{Eq:rate-isotropic-large-q0-spin}), (\ref{Eq:rate-isotropic-large-q0-spinless}), (\ref{Eq:relaxation-rate-2D-3D-final}), (\ref{Eq:FermiSpinfull}) and (\ref{Eq:FermiSpinless}).
At the largest energies,  $p_1\gg q_0$, the relaxation rate is given by Eq. (\ref{Eq:relaxation-rate-2D-3D-final})  and decays as $1/p_1$, irrespectively of the presence of spin.  In the intermediate momentum range, $\pf\ll p_1\ll q_0$, the relaxation rate for spinful (spinless) fermions grows as $p_1$ ($p_1^5$) [see Eqs. (\ref{Eq:rate-isotropic-large-q0-spin}) and (\ref{Eq:rate-isotropic-large-q0-spinless}), respectively]. Finally, in the Fermi-liquid regime, $p_1-\pf\ll \pf$, the relaxation rate exhibits universal $(p_1-\pf)^2$ scaling, with a prefactor that is smaller for spinless fermions 
due to Hartree-Fock cancellation, see Eqs. (\ref{Eq:FermiSpinfull}) and (\ref{Eq:FermiSpinless}).
}
\label{Fig:3DSchematic}
\end{figure}

The presence of spin becomes unimportant in the parameter regime  $\pf\lesssim q_0\ll p_1$ (including the physically most relevant case $q_0\sim \pf$). In this situation,
the interaction $w_{p_1/2}(0, \phi^\prime)$ is of the order of $V_0$ at $\phi^\prime =0$ but decays quickly beyond $\phi^\prime \sim q_0/p_1$, thus effectively limiting the available region of $\phi^\prime$. Thus, we get
\begin{equation}
\frac{1}{\tau_{p_1}}\sim m V_0^2 \frac{\pf^D q_0^{D-1}}{p_1}\,,\qquad   \pf\lesssim q_0\ll p_1. 
\label{Eq:relaxation-rate-2D-3D-final}
\end{equation}
Equation (\ref{Eq:relaxation-rate-2D-3D-final}) constitutes the main result of this section. It shows that
the relaxation rate of an ultra-hot particle exhibits a universal $1/p_1$ scaling in any spatial dimension $D\geq 2$ and irrespectively of the presence of spin.   

At low momenta, the relaxation rate $1/\tau_{p_1}$ follows the characteristic Fermi-liquid scaling, 
$1/\tau_{p_1}\propto (\pf-p_1)^2$ (up to logarithmic factors in $D=2$). Explicitly,  for spinful particles 
\begin{equation}
\frac{1}{\tau_{p_1}}\sim m V_0^2 \pf^{2D-4} (p_1-\pf)^2, \qquad p_1-\pf\ll \pf\lesssim q_0,
\label{Eq:FermiSpinfull}
\end{equation}
while for electrons without spin the prefactor is smaller because of the Hartree-Fock cancellation,
\begin{equation}
\frac{1}{\tau_{p_1}}\sim m V_0^2\frac{\pf^{2D} }{q_0^4}(p_1-\pf)^2, \qquad p_1-\pf\ll \pf\lesssim q_0.
\label{Eq:FermiSpinless}
\end{equation}

Combining Eqs. (\ref{Eq:rate-isotropic-large-q0-spin}) - (\ref{Eq:FermiSpinless}),
we conclude 
that the  relaxation rate generically exhibits a non-monotonic behavior, as  illustrated in Fig.  \ref{Fig:3DSchematic} for the case $D=3$. In this figure, dashed and solid lines correspond to spinless and spinful fermions, respectively.  After an initial increase at relatively low momenta, $p_1-\pf\ll \pf$ (which  follows the universal 
Fermi-liquid scaling but with different  prefactors for spinful and spinless cases) and an intermediate scaling regime at $\pf\ll p_1\ll q_0$, the relaxation rate starts to decrease  as $1/p_1$ independently of the presence of spin and of dimensionality of the system. 

\begin{figure}
\includegraphics[width=230pt]{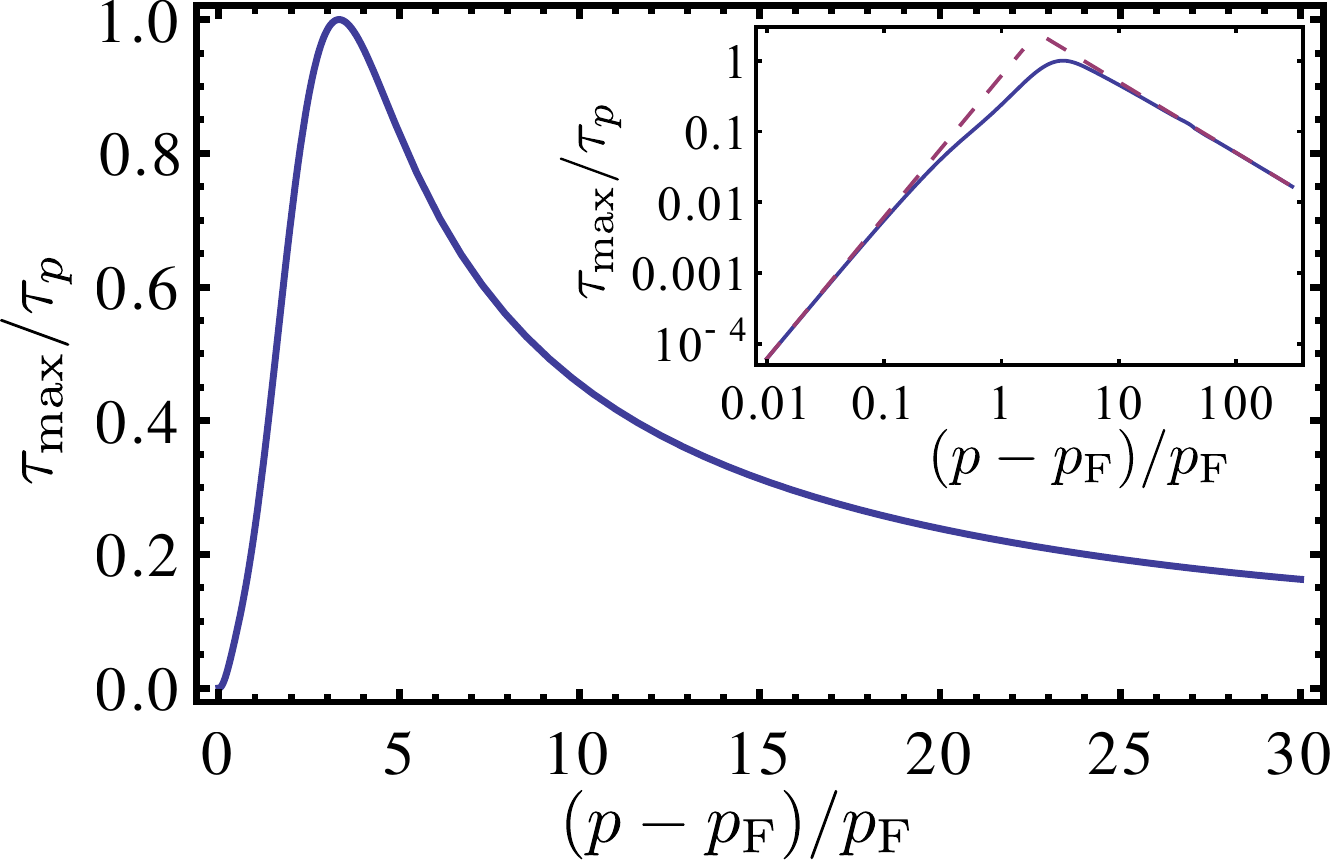}
\caption{Dependence of the relaxation rate on momentum $p_1$ [as given by  Eq. (\ref{Eq:relaxation-rate-2D-3D-convenient})]. Model interaction $V(q)=V_0 e^{-q^2/2\pf^2}$ and $D=3$ were used to generate the plot.
The relaxation rate exhibits a non-monotonic behavior with a maximum at $p_1\simeq 4\pf$ and was normalized by its maximal value $\tau_{\rm max}^{-1}$. Inset: the same dependence in the log-log scale. Dashed lines show the analytic predictions $1/\tau_p \propto (p-\pf)^2$ and $1/\tau_p \propto 1/p$ at $p-\pf \ll \pf$ and $p \gg \pf$, respectively. 
}
\label{Fig:Isotropic_rate}
\end{figure}

Our analytic predictions for the relaxation rate are in excellent agreement with the direct numerical evaluation of the integral (\ref{Eq:relaxation-rate-2D-3D-convenient}), as illustrated by Fig.~\ref{Fig:Isotropic_rate}. A model interaction $V(q)=V_0e^{-q^2/2\pf^2}$ was used to generate this plot. 

Concluding this section, let us discuss the applicability of our results to the case of Coulomb interaction between particles. To this end, we recall that the large-energy asymptotic  behavior  (\ref{Eq:relaxation-rate-2D-3D-final}) relies essentially on (i) the kinematics of the collision process and (ii) the inability of the interaction to transfer too large momenta [see discussion before Eq. (\ref{Eq:relaxation-rate-2D-3D-final})]. A careful analysis shows that despite a relatively slow decay of the screened Coulomb interaction, 
\begin{equation}
V_{\rm 2D}(q)=\frac{2\pi e^2}{\kappa+q}\,, \qquad V_{\rm 3D}(q)=\frac{4\pi e^2}{\kappa^2+q^2}
\label{screened-coulomb}
\end{equation}
(with $\kappa$ being the inverse screening radius), it is still fast enough for the scaling law 
(\ref{Eq:relaxation-rate-2D-3D-final}) to apply.  
More generally, the range of applicability of \eqref{Eq:relaxation-rate-2D-3D-final} extends  to interaction potentials $V(q)$ that decay at large momenta as $1/q^{\alpha}$  with $\alpha>(D-1)/2$. Indeed, inspecting Eqs.~\eqref{Eq:wDefSpin} and \eqref{Eq:Vpm}, we see that $w_q(0,\phi^{\prime})\sim q^{-2\alpha}\sin^{-2\alpha}(\phi^{\prime}/2)$. If $\alpha>(D-1)/2$, the integration in Eq.~\eqref{Eq:relaxation-rate-2D-3D-semifinal} is dominated by small $\phi^{\prime}\sim q_0/p_1$ and we obtain the scaling \eqref{Eq:relaxation-rate-2D-3D-final}. 
Using  $V_0 \sim e^2/\kappa$  in 2D, $V_0 \sim e^2/\kappa^2$ in 3D, and $q_0 \sim \kappa$, we get the relaxation rate of ultra-hot particles interacting via screened Coulomb interaction (\ref{screened-coulomb}) with $\kappa \gtrsim \pf$:
\begin{equation}
\frac{1}{\tau_{p_1}}\sim \frac{m e^4}{p_1}\times \left\{\begin{array}{cc}
\displaystyle \frac{\pf^2}{\kappa}, & \quad D=2,\\[0.4cm]
\displaystyle \frac{\pf^3}{\kappa^2}, & \quad D=3.
\end{array}\right.
\label{coulomb}
\end{equation}

The origin of the $1/p_1$ scaling (\ref{Eq:relaxation-rate-2D-3D-final}),  (\ref{coulomb}) of the  relaxation rate with the inverse momentum of the hot particle can be explained in the following way. 
In view of the suppression of the interaction at momentum transfers exceeding $q_0$, the energy transferred to a particle emerging form the Fermi sea can be at most of the order of $q_0^2/2m$. Such a change of the energy of a hot particle corresponds to a momentum transfer in the direction of $\bf p_1$ of the order of 
$q_0^2 /p_1 $, where we used the value $p_1/m$ of the hot-particle velocity.  This implies a reduction of the phase space by a factor $q_0/p_1 \ll 1$, explaining the $1/p_1$ scaling of $1/\tau$.  The specific $1/p_1$ form is thus related to the parabolicity of the spectrum. On the other hand, the decay of the relaxation rate with $p_1$ is more generic and will take place for any dispersion law with velocity increasing as a function of momentum.


\section{Multichannel quantum wires}
\label{Sec:MultiBand}

In the preceding section, we have presented a comprehensive analysis of the high-energy relaxation process for the particle with isotropic quadratic energy spectrum in $D=2$ and $3$ spatial dimensions. 
Let us now turn to the analysis of the relaxation in multichannel quantum wires and demonstrate that the non-monotonic behavior found in Sec. \ref{Sec:2D-3D}  persists also in this case. 

\subsection{Setup}
\label{Sec:Model}

\begin{figure}
\centering
\includegraphics[width=200pt]{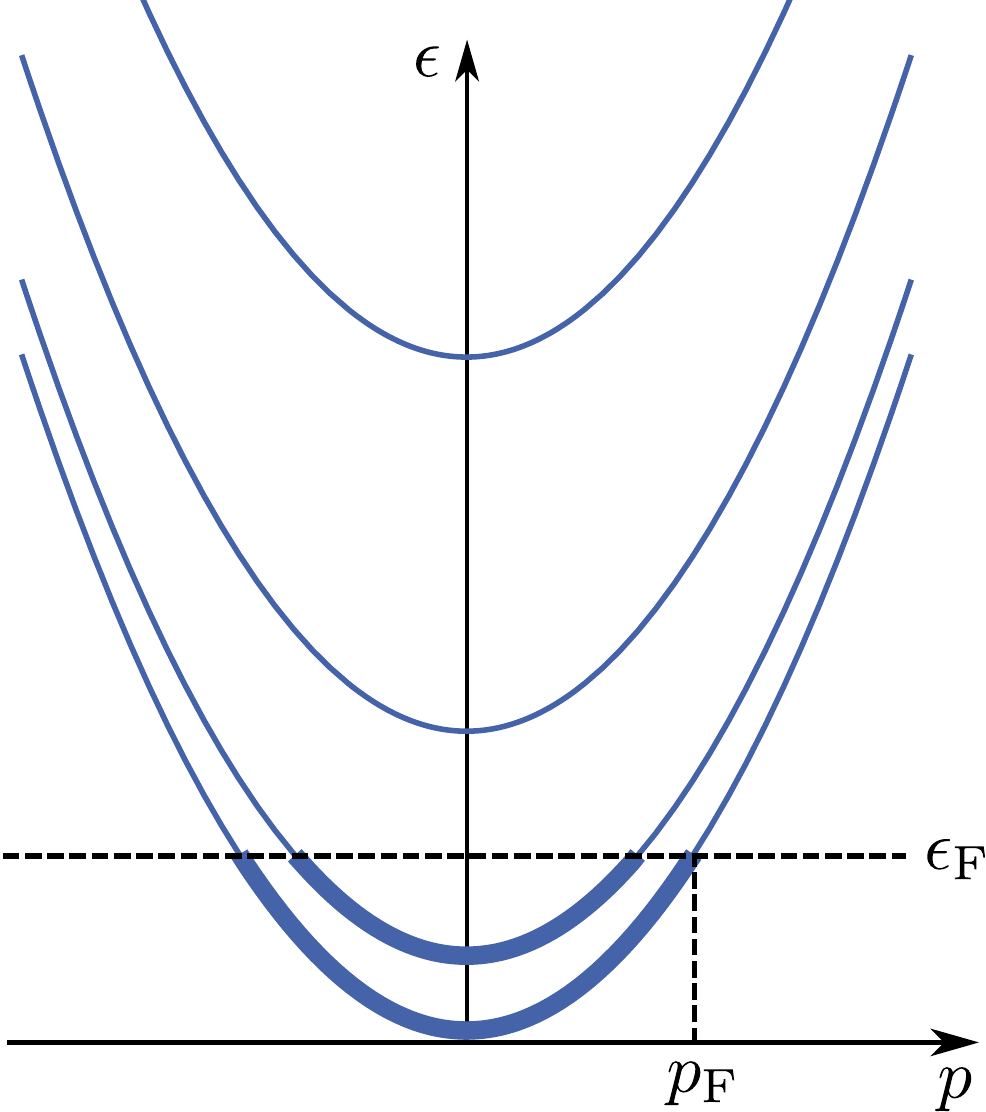}
\caption{1D bands of a quantum wire: energy $\epsilon$ as a function of the momentum $p$ along the wire. We model a multi-channel quantum wire by a collection of parabolic bands labeled by the index ${\bf n}$ of transverse quantization. The Fermi sea (bold sections of lines) occupies one or more low-lying bands; $\pf$ denotes the Fermi momentum in the lowest band. }
\label{Fig:BandStructuture}
\end{figure}

We begin by formulating the model. A quasi-1D wire that we will consider hosts 1D energy  bands with quadratic dispersion enumerated by an index ${\bf n}$ that corresponds to the transverse quantization,
\begin{equation}
\epsilon_{\bf n}(p)=\frac{p^2}{2m}+\Delta_{\bf n}.
\label{Eq:epsilonN}
\end{equation}   
Here, $p$ is the momentum in the direction along the wire (which we choose as $z$-axis) and $\Delta_{\bf n}$ sets the bottom of the ${\bf n}$-th band. The Fermi see resides in one or more of the low-lying bands. We denote by $\pf$ the Fermi momentum in the lowest band, see Fig. \ref{Fig:BandStructuture}.

The electrons populating the wire interact via an interaction $V(|{\bf r}|)=V(|{\bf r}^{\perp}|, z)$ that translates, in the band picture, into
\begin{equation}
\begin{split}
V_{{\bf n}_1, {\bf n}_2; {\bf n}_1^\prime {\bf n}_2^\prime}(q)\equiv&\\
\int d{\bf r}_{1}^{\perp}d{\bf r}_{2}^{\perp}V(|{\bf r}^{\perp}_{1}-&{\bf r}^{\perp}_{2}|, q) \psi_{{\bf  n}_1}^*({\bf r}^{\perp}_{1})
\psi_{{\bf n}_2}^*({\bf r}^{\perp}_{2}) \psi_{{\bf  n}_1^\prime}({\bf r}^{\perp}_{1})
\psi_{{\bf n}_2^\prime}({\bf r}^{\perp}_{2}).
\end{split}
\end{equation}
Here the integration runs over the cross-section of the wire and $\psi_{\bf n}({\bf r}^{\perp})$ are wave functions of transversal quantization. 

\begin{figure}
\centering
\includegraphics[width=220pt]{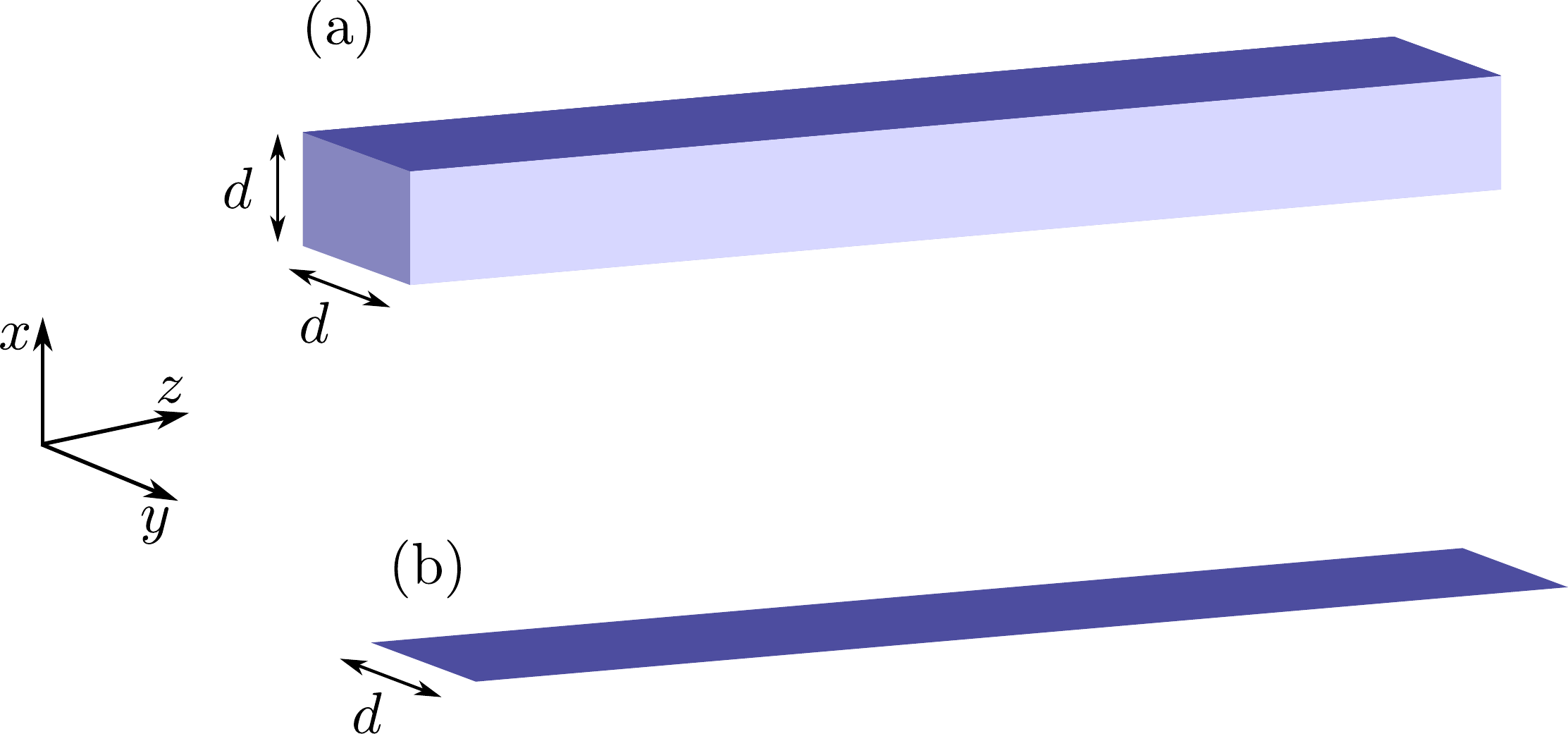}
\caption{Quasi-1D geometry. (a) Two finite dimensions of size $d$ to study the crossover to 3D. (b) One finite dimension to study the crossover to 2D.\label{Fig:geometry}}
\label{geometry}
\end{figure}

Our goal is to study the relaxation of a hot  electron injected at some momentum $p_1$  into  one of the bands ${\bf n}_1$ such that its energy $\epsilon_{{\bf n}_1}(p_1)\gg \epsilon_F$.  In order to accomplish this task, one needs to know the band positions $\Delta_{\bf n}$ and the transverse-quantization wave functions  $\psi_{\bf n}({\bf r}^{\perp})$.  As we discuss below,
the high-energy scattering processes involve  excitations of electrons into high energy bands. We thus expect that the corresponding transversal wave functions are largely independent of microscopic details and can be approximated by plane waves (e.g., with periodic boundary conditions). In such a setting, the index $\bf n$ can be identified with the transversal momentum. Its precise nature depends on the transversal dimensionality of the wire.
In this work, we consider two physically relevant cases (see Fig. \ref{geometry}): (i) a wire  defined as a stripe of width $d$ in a 2D electron gas,  in which case $\bf{n}$ is just an integer number, ${\bf n}=0, \pm 1, \ldots$, and (ii)  a wire with 2D cross-section (of characteristic size $d$), in which case ${\bf n}=(n_x, n_y)$ becomes a 2D integer vector. 

In accordance with our identification of the band index with the transversal momentum, we choose 
\begin{equation}
\Delta_{\bf n}=\Delta_0|{\bf n}|^2\,, \qquad \Delta_0\sim \frac{1}{m d^2},
\end{equation}
and impose the momentum conservation condition on the matrix elements of the interaction
\begin{equation}
V_{{\bf n}_1, {\bf n}_2; {\bf n}_1^\prime {\bf n}_2^\prime}(q)=\frac{1}{d^{D_\perp}}V_{{\bf n}_1-{\bf n}^\prime_1}(q)
\delta_{{\bf n}_1+{\bf n}_2-{\bf n}^\prime_1-{\bf n}^\prime_2}.
\label{matrix-element-quasi-1D}
\end{equation}
Here, $D_\perp=1, 2$ is the transversal dimensionality of the wire. In analogy with Sec.~\ref{Sec:2D-3D},
we assume the interaction to be approximately isotropic, 
$V_{{\bf n}_1-{\bf n}^\prime_1}(q)\simeq V\left(Q\equiv \sqrt{q^2+4\pi^2|{\bf n}|^2/d^2}\right)$,
and characterized by a single momentum scale $q_0$. Specifically, the interaction $V(Q)$ can be expanded at not too large momenta,
\begin{equation}
V(Q) \simeq V_0\left(1-Q^2/q_0^2\right)\,, \qquad Q\ll q_0,
\label{Eq:interaction-Q1D}
\end{equation}
and decays fast enough at $Q \gg q_0$. Physically, this means the relaxation rate is determined by momentum transfers less than or of the order of $q_0$.



\subsection{Q1D setup with one lateral dimension}
\label{Sec:Q1D-1-lateral}

We start our analysis of quantum wires by considering the case of a wire realized as a stripe of width $d$ in a 2D electron gas. Such a model can be viewed as describing the crossover from one to two spatial dimensions.  Assuming a weak interaction, we employ the golden rule to calculate the relaxation rate: 
\begin{eqnarray}
\frac{1}{\tau} &=& \frac{1}{2!}\int \diff p_2\, \diff \pp{1}\, \diff \pp{2}\!\! \!\!\sum_{n_2,\np{1},\np{2}}\delta(p_1+p_2-\pp{1}-\pp{2})
\nonumber
\\
& \times & \delta_{n_1+n_2,\np{1}+\np{2}}\delta(\epsilon_1+\epsilon_2-\epsilon_{1^{\prime}}-\epsilon_{2^{\prime}})   \nonumber \\
& \times & F(\lambda_1,\lambda_2;\lambda_{1^{\prime}},\lambda_{2^{\prime}}).
\label{Eq:relaxation-rate-1-lateral}
\end{eqnarray}
Here we  have introduced  a shorthand notation $\lambda_i=(n_i,p_i)$, and $\epsilon_i \equiv \epsilon_{n_i}(p_i)$ is given by Eq.~\eqref{Eq:epsilonN}. The summation over transversal momentum (that is just a scalar integer in the present case) is restricted by the conservation law assumed in our model. The function $F$ in Eq.~(\ref{Eq:relaxation-rate-1-lateral}) reads
\begin{eqnarray}
F(\lambda_1,\lambda_2;\lambda_{1^{\prime}},\lambda_{2^{\prime}}) &=& \nf(\epsilon_2) [1-\nf(\epsilon_{1^{\prime}})][1-\nf(\epsilon_{2^{\prime}})]
\nonumber \\
&\times &  \left|M_{\lambda_1,\lambda_2}^{\lambda_{1^{\prime}},\lambda_{2^{\prime}}}\right|^2,
\end{eqnarray}
with the modulus squared of the matrix element given by
\begin{equation}
\left|M_{\lambda_1,\lambda_2}^{\lambda_{1^{\prime}},\lambda_{2^{\prime}}}\right|^2=\frac{1}{d^2}\left[V_{n_1-\np{1}}(p_1-\pp{1})-V_{n_1-\np{2}}(p_1-\pp{2})\right]^2
\label{matrix-element-no-spin}
\end{equation}
or 
\begin{eqnarray}
\left|M_{\lambda_1,\lambda_2}^{\lambda_{1^{\prime}},\lambda_{2^{\prime}}}\right|^2 &=& \frac{1}{d^2}\Bigl[\left[V_{n_1-\np{1}}(p_1-\pp{1})\right]^2
\nonumber \\
&+& \left[V_{n_1-\np{2}}(p_1-\pp{2})\right]^2 \nonumber \\
&-& V_{n_1-\np{1}}(p_1-\pp{1})V_{n_1-\np{2}}(p_1-\pp{2})\Bigr]
\end{eqnarray}
in the absence and presence of spin, respectively. 
It proves convenient to introduce the (longitudinal) momentum transfer $q=p_1-p_1^\prime$ as one of the integration variables.  The integrations over $p_2$ and $\pp{1}$ can be performed with the help of the $\delta$ functions describing the conservation of energy and longitudinal momentum, yielding
\begin{eqnarray}
\frac{1}{\tau} & = &\frac{m}{2}\sum_{\{n_i\}}\int \frac{\diff q}{|q|} \nonumber
\\
&\times & F_{\{n_i\}}\left(p_1,p_1-q+\frac{m\Delta_{\mathrm{eff}}}{q};p_1-q,p_1+\frac{m\Delta_{\mathrm{eff}}}{q}\right),
\nonumber \\ &&
\label{Eq:relaxation-Q1D-1-lateral-integration-over-q}
\end{eqnarray}
where we introduced the energy 
\begin{equation}
\Delta_{\mathrm{eff}}=\Delta_0(n_1^2+n_2^2-\np{1}^2-\np{2}^2).
\label{Delta-eff}
\end{equation}

\begin{figure}
\includegraphics[scale=0.35]{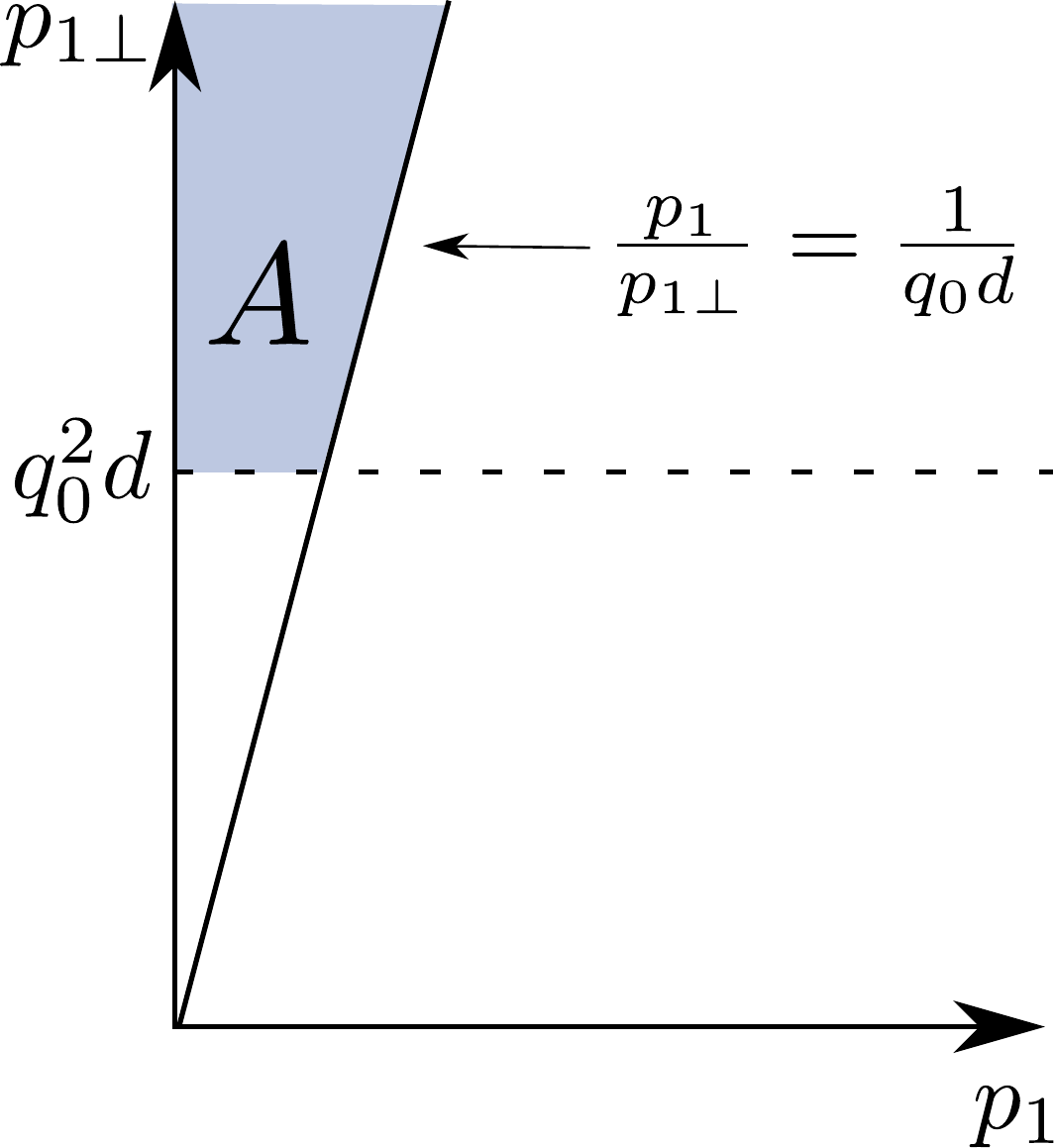}
\caption{Non-universal regime  (\ref{region-A}) in the plane spanned by the longitudinal $p_1$ and transversal $p_{1\perp}$ momentum in quasi-1D wires. In this regime, the relaxation rate decays as a function of energy $\epsilon$ faster than $1/\sqrt{\epsilon}$. The actual dependence is non-universal since it depends on the large-momentum tail of $V(q)$.}
\label{Fig:non-universal-regime}
\end{figure}

A detailed analysis of this expression can be found in Appendix \ref{App:Q1D-1-lateral}.
We find the following results for the relaxation rate of an ultra-hot electron with the total momentum $p_1^{\rm tot} = \sqrt{p_1^2 + p_{1\perp}^2} \gg \pf$  in a quasi-1D wire constituting a 2D strip:
\begin{equation}
\frac{1}{\tau_{\epsilon}}\sim
\left\{ \begin{array}{ll}
 m V_0^2\pf^2, & \quad \pf \ll p_1^{\rm tot} \ll q_0, \ \ \text{spinful}, \\[0.3cm]
\displaystyle m^3 V_0^2\frac{\pf^2}{q_0^4}\epsilon^2, & \quad \pf \ll p_1^{\rm tot} \ll q_0, \ \  \text{spinless}, \\[0.3cm]
\displaystyle  \sqrt{m}V_0^2\pf^2\frac{q_0}{\sqrt{\epsilon}}, & \quad \pf\lesssim q_0 \ll p_1^{\rm tot}; \, (p_{1\perp},p_{1}) \not \in  A,
 \end{array} \right.
 \label{wires-2D-results}
 \end{equation}
where we introduced the transversal momentum $p_{1\perp}=n_1/d$. Equation (\ref{wires-2D-results}) largely coincides 
with the corresponding results for an isotropic 2D system, Sec.~\ref{Sec:2D-3D}. The difference to the 2D situation occurs when the momentum is high and is almost in the transversal direction. More specifically, if the momentum $(p_{1\perp},p_1)$ belongs to the region defined by two equations (see Fig.~\ref{Fig:non-universal-regime}),
\begin{equation}
\text {region A:} \qquad p_{1\perp}>q_0^2d \quad \mathrm{and} \quad \frac{p_{1}}{p_{1\perp}}< \frac{1}{q_0 d},
\label{region-A}
\end{equation}
the relaxation rate is additionally suppressed in comparison with the 2D result due to transverse energy quantization. To understand the reason for this, let us consider the case of an initial momentum pointing out exactly in the transverse direction, $p_1 = 0$, which corresponds to the bottom of a certain high energy band. A decay process then necessarily involves a transition to a lower energy band, which means a momentum transfer $\gtrsim 1/d$  in the direction of the initial momentum. If this momentum transfer is larger than $q_0^2/p_1$, such a process will be parametrically suppressed, as explained  in the end of Sec.~\ref{Sec:2D-3D}. This yields the first of the conditions (\ref{region-A}). A more detailed analysis shows that this suppression happens not only at $p_1=0$ but also in a range of $p_1$ given by the second condition (\ref{region-A}).  The relaxation rate in the regime (\ref{region-A}) is determined by 
the large-momentum tail of the interaction, which leads to a faster decay of the relaxation rate than $1/\sqrt{\epsilon}$. The actual form of the decay is non-universal as it depends on the large momentum behavior of $V(q)$.  In the continuum limit (fixed momenta $\pf$, $q_0$, $p_{1}$, $p_{1\perp}$ and $d\to \infty$), the regime (\ref{region-A}) disappears and we recover the usual isotropic 2D result.

As in an isotropic 2D system, the above results apply also for interaction potentials that decay at large $q$ as a power law $1/q^\alpha$ with $\alpha > 1/2$. This includes, in particular, the case of a (screened) Coulomb interaction $V_{\rm 2D}(q)$,  Eq.~(\ref{screened-coulomb}).


\subsection{Q1D setup with two lateral dimensions}
\label{Sec:Q1D-2-lateral}

After the discussion of the setup with one finite transversal dimension, we now turn to the case of a 3D wire, i.e., of a quasi-1D setup with two lateral dimensions. In comparison to the discussion of Sec.~\ref{Sec:Q1D-1-lateral}, we need to associate two discrete indices with each electron state: $n_i \to \nb_i=(n_{i,x},n_{i,y})$, and replace in the matrix element $V^{(2)}_{n}(p)/d$ by $V^{(3)}_{\nb}(p)/d^2$, see Eq.~\eqref{matrix-element-quasi-1D}.

In comparison to the situation with only one lateral dimension, there is an additional subtlety here. If the integer vectors $(\nb_{1^{\prime}}-\nb_1)$ and $(\nb_{1^{\prime}}-\nb_2)$ are perpendicular to each other, the longitudinal momentum transfer $q=0$ is allowed (``vertical relaxation"). According to Eq.~\eqref{Eq:relaxation-Q1D-1-lateral-integration-over-q}, the contributions from those processes to the relaxation rate are formally divergent in a logarithmic fashion. At zero temperature, such processes are only allowed if the longitudinal momentum $p_1$ is smaller than the Fermi momentum $p_{\mathrm{F}}$. The logarithmic singularity gets regularized due to broadening of the $\delta$ function related to energy conservation by other processes. 
As a result, the contribution of such processes gets an additional logarithmic factor $\sim \ln (p_{1\perp}/\pf)$  in comparison with ``non-singular'' processes.
On the other hand, because of the above orthogonality condition of the vectors $(\nb_{1^{\prime}}-\nb_1)$ and $(\nb_{1^{\prime}}-\nb_2)$, only a small fraction of all processes belong to the ``singular'' class. This results in a suppression of ``singular'' processes by a factor $(\pf q_0d^2)^{-1} \ll 1$. 
In a parametrically broad range of momenta $p_{1\perp}$ of the hot electron, this power-law suppression will be more important than the logarithmic enhancement, so that the ``singular'' processes will yield a subleading contribution. Only at very high $p_{1\perp}$ will the singular processes give a dominant contribution (assuming one can still apply the model at such energies). This will, however, modify our conclusion on $1/p^{\rm tot}_1$ decay of the relaxation rate  only by an additional logarithmic factor.   
Furthermore, ``singular'' processes are absent in more realistic models in which the energy dependence deviates from perfect parabolicity. In the following we thus discard this subclass of processes. 

Analyzing ``non-singular'' processes, we find for the relaxation rate of an ultra-hot electron with the total momentum $p_1^{\rm tot} = \sqrt{p_1^2 + p_{1\perp}^2} \gg \pf$ in a 3D wire (see Appendix~\ref{App:Q1D-2-lateral} for detail) 
\begin{equation}
\frac{1}{\tau_{\epsilon}}\sim
\left\{ \begin{array}{ll}
m^{3/2} V_0^2 \pf^3 \sqrt{\epsilon}, & \qquad \pf \ll p_1^{\rm tot} \ll q_0, \ \ \text{spinful}, \\[0.3cm]
\displaystyle m^{7/2} V_0^2 \frac{\pf^3}{q_0^4} \epsilon^{5/2}, & \qquad \pf \ll p_1^{\rm tot} \ll q_0, \ \  \text{spinless}, \\[0.3cm]
\displaystyle \sqrt{m}V_0^2 \pf^3 \frac{q_0^2}{\sqrt{\epsilon}}, & \qquad \pf\lesssim q_0 \ll p_1^{\rm tot}, \\
& \qquad  (\mathbf{p}_{1\perp},p_1) \not \in A,
 \end{array} \right.
 \label{wires-3D-results}
 \end{equation}

in consistency with the corresponding results for an isotropic 3D system, Sec.~\ref{Sec:2D-3D}.  In full analogy to the case of a wire constituting a 2D strip, there is the regime A, Eq.~\eqref{region-A} where the discreteness of the transverse spectrum leads to an additional suppression of the relaxation rate. 
 
As in an isotropic 3D system, these results apply also for interaction potentials that decay at large $q$ as a power law $1/q^\alpha$ with $\alpha > 1$. This includes, in particular, the case of a (screened) Coulomb interaction $V_{\rm 3D}(q)$,  Eq.~(\ref{screened-coulomb}).


 \section{1D wires  and triple collisions}
 \label{Sec:1D}
 
 In the previous sections, we have shown that non-monotonicity  is a generic feature of the 
  energy dependence  of the relaxation rate  in $D\geq 2$ spatial dimensions as well in quasi-1D wires hosting many subbands. In the present section, we consider the case of a single-channel 1D wire with parabolic dispersion.
 
 In one dimension, energy and momentum conservation restrict the two-particle collisions studied in Sec. \ref{Sec:2D-3D} to permutations of the momenta of colliding particles. Thus, two-particle collisions can not lead to relaxation and one has to study three-particle collision processes. In the ``low-energy'' domain (energy much smaller than the Fermi energy), such a study was accomplished in a number of works\cite{Lunde07,Khodas2007,imambekov09,imambekov11,micklitz11,ristivojevic13,apostolov13,Lin2013,MatveevFurusaki2013,PGM2013,Matveev2013,Dmitriev2012,Protopopov2014,Gangardt2014,protopopov13}. Little is known, however, about the ultra-high-energy case $\epsilon\gg \epsilon_F$. This gap is filled in the present section.

In the case of triple collisions, the Fermi golden rule  reads [cf. Eq. (\ref{Eq:relaxation-rate-2D-3D}) and notations therein]:
\begin{widetext}
\begin{equation}
\frac{1}{\tau_{p_1}}=\frac{1}{2!3!}\int  dp_2 dp_3 dp_1^\prime dp_2^\prime  dp_3^\prime \,\delta\left(E_{\rm i}-E_{\rm f}\right)
\delta\left(P_i-P_f\right) \nf(\epsilon_2) \nf(\epsilon_3)[1-\nf(\epsilon_{1}^{\prime})][1-\nf(\epsilon_{2}^{\prime})][1-\nf(\epsilon_{3}^{\prime})]
\left|M_{p_1,p_2, p_3}^{p_1^\prime, p_2^\prime, p_3^\prime}\right|^2.
\label{Eq:relaxation-rate-1D}
\end{equation}
\end{widetext} 
Here, $M_{p_1,p_2, p_3}^{p_1^\prime, p_2^\prime, p_3^\prime}$ is the matrix element for the triple collisions whose precise form will be discussed below. If the fermions have spin, the symbol $\left|M_{p_1,p_2, p_3}^{p_1^\prime, p_2^\prime, p_3^\prime}\right|^2$ implicitly incorporates the summation over spin indices. 

The energy and momentum conservation constraints are conveniently  resolved by parametrizing the momenta of colliding particles according to\cite{ristivojevic13, Protopopov2014} 
\begin{eqnarray}
p_k=\frac{P}{3}+q\cos\left[\phi+\frac{2\pi (k-1)}{3}\right]&\,, \quad k=1, 2, 3,\;
\label{Eq:1DParam1}
\\
p_k^\prime=\frac{P}{3}+q\cos\left[\phi^\prime+\frac{2\pi (k-1)}{3}\right]&\,, \quad k=1, 2, 3.\;
\label{Eq:1DParam2}
\end{eqnarray}
The variable $-\infty<P<\infty $ is nothing but the total momentum of the particles while $q\geq0$ is analogous to the 
relative momenta in the two-particle collision processes and fixes the total energy 
\begin{equation}
E_i=E_f=\frac{P^2}{6m}+\frac{3q^2}{4m}.
\end{equation}
  
One can now rewrite the collision rate (\ref{Eq:relaxation-rate-1D}) in the form [cf. Eq. (\ref{Eq:relaxation-rate-2D-3D-convenient}) of  Sec. \ref{Sec:2D-3D}]
\begin{widetext}
\begin{equation}
\begin{split}
\frac{1}{\tau_{p_1}}=\frac{m}{2}\int_{-\infty}^{\infty} dP \int^{\infty}_{\left|P/3-p_1\right|}q\, dq
\int_{-\pi}^\pi d\phi\, d\phi^\prime \;\delta\left[p_1 -p_1(P, q, \phi)\right]  
\nf(\epsilon_2) \nf(\epsilon_3)[1-\nf(\epsilon_{1}^{\prime})][1-\nf(\epsilon_{2}^{\prime})][1-\nf(\epsilon_{3}^{\prime})]
w_q(\phi, \phi^\prime).
\end{split}
\label{Eq:relaxation-rate-1D-convenient}
\end{equation}
\end{widetext}
Here, $p_1(P, q, \phi)$ is given by Eq. (\ref{Eq:1DParam1}) and $w_q(\phi, \phi^\prime)$ is the modulus squared of the  triple-collision matrix element, $\left|M_{p_1,p_2, p_3}^{p_1^\prime, p_2^\prime, p_3^\prime}\right|^2$,  evaluated on the mass shell. 

\begin{figure}
\includegraphics[width=230pt]{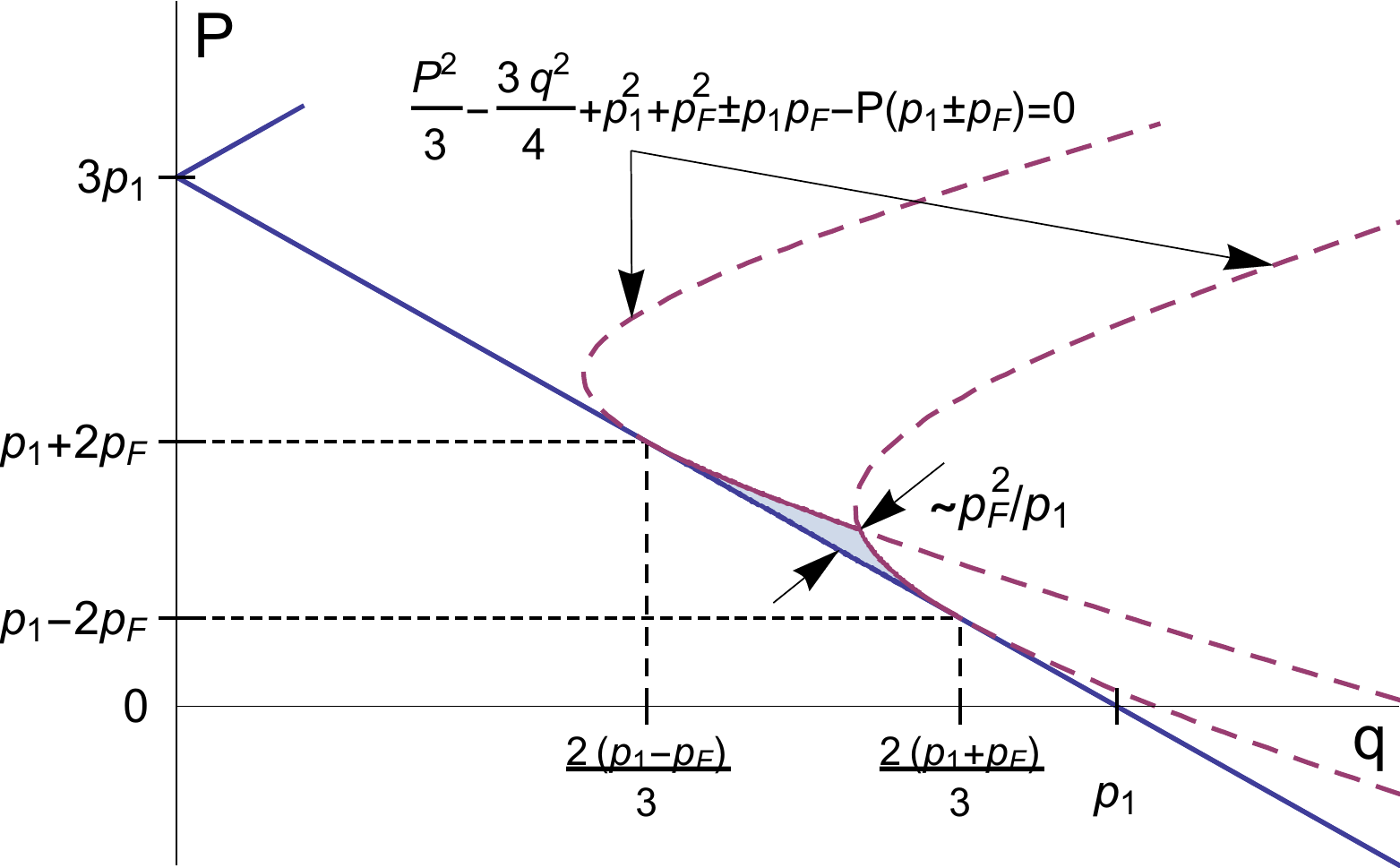}
\caption{Kinematics of 1D three-particle  scattering process in the $(q,P)$ plane. The Fermi distributions $\nf(\epsilon_2)$ and $\nf(\epsilon_3)$ restrict the allowed values of $q$ and $P$ to a vicinity of the point $(2p_1/3, p_1)$ represented by the shaded region between the line $P/3+q=p_1$ and two hyperbolas.  }
\label{Fig:PQPlane1D}
\end{figure}

The analysis of Eq. (\ref{Eq:relaxation-rate-1D-convenient}) proceeds to large extent along the same lines as that of Sec. \ref{Sec:2D-3D}. First of all, for $p_1\gg \pf$ the Fermi functions $\nf(\epsilon_2)$ and $\nf(\epsilon_3)$ restrict the 
integration in the $(q, P)$ plane to a small neighborhood of the point $(q=2p_1/3, P=p_1)$, see Fig. \ref{Fig:PQPlane1D}. 
The $\delta$ function fixes then the angle $\phi$ to
\begin{equation}
\phi_0=\pm\arccos \frac{p_1-P/3}{q}, \qquad |\phi_0|\lesssim\frac{\pf}{p_1}.
\label{Eq:phi0-1D}
\end{equation}
On the other hand, just as in the higher-dimensional situations and in contrast to the low-energy scattering, the Fermi factors associated to the outgoing particles are not important (provided that the interaction is capable of transferring momenta of the order of $\pf$, which is what we have assumed).
As a result, Eq. (\ref{Eq:relaxation-rate-1D-convenient})  reduces to 
\begin{equation}
\frac{1}{\tau_{p_1}}\sim m \pf^2\int_{-\pi}^{\pi}d\phi^\prime w_{2p_1/3}\left(\phi\sim \frac{\pf}{p_1}, \phi^\prime\right).
\label{Eq:rate1DSemiFinal}
\end{equation}

Equation (\ref{Eq:rate1DSemiFinal}) relies essentially only on the kinematics of the three-particle collision process. To complete our analysis, a dynamical input is necessary. The matrix element $M_{p_1,p_2, p_3}^{p_1^\prime, p_2^\prime, p_3^\prime}$ governing the dynamics arises in the second order of the perturbation theory in the two-body interaction\cite{Khodas2007, ristivojevic13, MatveevFurusaki2013, Dmitriev2012}.  
In the spinless case it is given by the vacuum expectation value 
\begin{equation}
M_{p_1,p_2, p_3}^{p_1^\prime, p_2^\prime, p_3^\prime}=
\left\langle a_{p_3}a_{p_2}a_{p_1}\right| \hat V\frac{1}{E-\hat H_0+i0}\hat V\left| a^{\dagger}_{p_1}a^{\dagger}_{p_2}a^{\dagger}_{p_3}\right\rangle,
\label{Eq:MTrippleBasic}
\end{equation}
where $\hat H_0$ and $\hat V$ are free and interaction parts of the Hamiltonian, respectively.
A straightforward albeit somewhat lengthy algebra allows one to rewrite Eq. (\ref{Eq:MTrippleBasic}) as (see Appendix \ref{App:1D-amplitude})
\begin{widetext}
\begin{equation}
M_{p_1,p_2, p_3}^{p_1^\prime, p_2^\prime, p_3^\prime}=M_{q}(\phi, \phi^\prime)=
\frac{4m}{3q^2}\frac{\sum_{j, k=0}^{2}\Gamma_q\left(\phi+\frac{2\pi k}{3}, \phi^\prime+\frac{2\pi j}{3}\right)\left[
\cos\left(\phi+\frac{2\pi k}{3}\right)-\cos\left(\phi^\prime+\frac{2\pi j}{3}\right)
\right]}{\cos 3\phi-\cos3\phi^\prime},
\label{Eq:MTrippleFinal}
\end{equation}
\end{widetext}
with 
\begin{equation}
\begin{split}
\Gamma_q(\phi, \phi^\prime)=
[V(q_{10})-V(q_{20})][V(q_{01})-V(q_{02})]\\-2V(q_{12})V(q_{21})+2V(q_{11})V(q_{22}),
\end{split}
\end{equation}
where we have introduced shorthand notations for various momentum transfers involved in the process:
\begin{equation}
q_{kj}=q\left[\cos\left(\phi+\frac{2\pi k}{3}\right)-\cos\left(\phi^{\prime}+\frac{2\pi j}{3}\right)\right].
\end{equation}

The matrix element $M_q(\phi, \phi^\prime)$ is an odd periodic function of each of the angles  with the period $2\pi/3$ reflecting the indistinguishability of the particles. 
The factor $1/q^2$ in Eq. (\ref{Eq:MTrippleFinal}) stems  from  the typical scaling of the energy denominators in Eq. (\ref{Eq:MTrippleBasic}), while the denominator $\cos3\phi-\cos3\phi^\prime$ takes into account the possibility for them to become small for particular configurations of angles $\phi=\phi^\prime \mod 2\pi/3$. Using the symmetry of the function $\Gamma_q(\phi, \phi^\prime)$ with respect to the exchange of $\phi$ and $\phi^\prime$, it is easy to see, however, that the vanishing of the denominator  in Eq. (\ref{Eq:MTrippleFinal}) does not  lead to a pole and the matrix element $M_{q}(\phi, \phi^\prime)$ 
is in fact a smooth function of angles. 

A further examination of Eq. (\ref{Eq:MTrippleFinal}) shows that, for $q\gg q_0\gtrsim \pf$, the function 
$M_{q}(\phi\sim \pf/q, \phi^\prime)$ is of the order of  $m \pf V_0^2/q_0q^2$ for $\phi^\prime \lesssim q_0/q$ and is strongly suppressed beyond this range of angles (up to aforementioned periodicity). More specifically, at fixed $\tilde{\phi}$ and $\tilde{\phi}^\prime$ and $q\gg q_0$ the matrix element possesses the scaling
\begin{eqnarray}
&& \hspace*{-1cm} M_{q}\left(q_0\tilde{\phi}/q,  q_0\tilde{\phi}^\prime /q \right) \nonumber \\
 & = & \frac{4m}{9q^2} \left[V^2(q_-)+V(0)(V(q_-)-q_{-}V^\prime(q_-))\right] \nonumber \\
&-& \langle q_-\rightarrow q_+ \rangle,
\label{Eq:MSpinlessScaling}
\end{eqnarray}
where $q_\pm=\sqrt{3}q_0(\tilde{\phi}\pm\tilde{\phi}^\prime)/2$ and $V^\prime$ stands for the derivative of interaction with respect to momentum. In deriving Eq. (\ref{Eq:MSpinlessScaling}), we have neglected all the terms with momentum transfer of order $q$. According to Eq.~(\ref{Eq:MSpinlessScaling}),  the $1/q^2$ scaling of the matrix element readily seen in Eq. (\ref{Eq:MTrippleFinal}) for generic values of angles  $\phi$ and $\phi^\prime$ survives also in the regime $\phi, \phi^\prime \lesssim  q_0/q$ where the energy denominators in Eq. (\ref{Eq:MTrippleFinal}) are of the order of $q_0^2/m$ only. The reason for this is an intricate cancellation between various interaction processes in Eq. (\ref{Eq:MTrippleFinal}).

For $\tilde{\phi}, \tilde{\phi}^\prime\lesssim 1$,  Eq.(\ref{Eq:MSpinlessScaling}) takes the form 
\begin{equation}
M_{q}\left(q_0\tilde{\phi}/q,  q_0\tilde{\phi}^\prime /q \right)\sim \frac{m V_0^2}{q^2}\tilde{\phi}\tilde{\phi}^\prime =   \frac{m V_0^2}{q_0^2}\phi\phi^\prime.
\label{Eq:MSpinlessScalingSmall}
\end{equation}
Setting $\tilde{\phi} \sim \pf/q_0\lesssim 1$ in  Eq. (\ref{Eq:MSpinlessScalingSmall}) and combining the resulting estimate with Eq. (\ref{Eq:rate1DSemiFinal}), we arrive at
\begin{equation}
\frac{1}{\tau_{p_1}}\sim \frac{m^3 \pf^4 V_0^4}{q_0 p_1^5} , \qquad p_1\gg q_0\gtrsim \pf \quad \text{(spinless)}.
\label{Eq:1DspinlessFinal}
\end{equation}

Equation (\ref{Eq:1DspinlessFinal}) constitutes one of the main results of this section. It shows that, for 
spinless fermions interacting via an interaction with characteristic momentum transfer larger than $\pf$, the relaxation rate at ultrahigh energies, $p_1\gg q_0$, is suppressed in a power-law fashion as $1/\tau_p \propto p^{-5}$. Let us note a qualitative similarity of this result with the higher dimensional analog, Eq. (\ref{Eq:relaxation-rate-2D-3D-final}). In both cases the suppression of the available phase space by the maximal momentum that can be transferred by the interaction brings the factor $1/p_1$. In addition, in the case of triple collisions  that govern the relaxation in 1D systems a factor $1/p_1^4$ arises from the partial cancellations of direct and exchange terms in the three-particle matrix element.

In the intermediate regime where the external momentum $p_1$ is smaller than the momentum scale $q_0$ but still much larger than the Fermi momentum, we expand the interaction potential $V(q)$ to fourth order. We obtain for the squared matrix element
\begin{equation}
w_q(\phi,\phi^{\prime})\sim\frac{m^2V_0^4q^{12}\sin^2(3\phi)\sin^2(3\phi^{\prime})}{q_0^{16}}.
\label{Eq:1D-squared-matrix-element-large-q0}
\end{equation}
After the integration over $\phi^{\prime}$ and making use of the estimates $q\sim 2p_1/3$ and $\phi \sim \pf/p_1$, we find the scaling
\begin{equation}
\frac{1}{\tau_{p_1}}\sim \frac{m^3 \pf^4 V_0^4}{q_0^{16}}p_1^{10}, \qquad \pf \ll p_1 \ll q_0 \quad \text{(spinless)}.
\label{Eq:1Dspinless_large_q0}
\end{equation}
In the low-energy regime we find the scaling (see Appendix \ref{App:1D-low-energy} for details)  
\begin{equation}
\begin{split}
\frac{1}{\tau_{p_1}}\sim \frac{m^3 \pf^6 V_0^4}{q_0^{16}}(p_1-\pf)^8, \quad p_1-\pf\ll &\pf\ll q_0, 
\\
&(\mathrm{spinless}),
\end{split}
\label{Eq:1Dspinless_low_p1}
\end{equation}
that coincides with the behavior found in Ref. \onlinecite{Khodas2007} if the Fermi momentum is of the same order as the momentum scale of the interaction $q_0$.

For fermions with spin, the matrix element is given by an analog of Eq. \eqref{Eq:MTrippleBasic}, see Appendix \ref{App:1D-amplitude}. After the spin summation and the parametrization according to Eqs. \eqref{Eq:1DParam1} and \eqref{Eq:1DParam2}, the square of the matrix element assumes a form that is similar to Eq. \eqref{Eq:MTrippleFinal}. In contrast to the spinless case, non-integrable poles appear in the squared matrix element for $\phi^{\prime}=\phi \mod 2\pi/3$. However, at zero temperature the Fermi functions for final particles restrict the integration over the angle $\phi^{\prime}$ to the region that excludes those poles with the typical width of the excluded region of the order $\pf/p_1$.  At non-zero temperature, the expression for the matrix element has to be regularized. As shown in Ref. \onlinecite{Dmitriev2012}, the divergence is related to two consecutive two-particle scattering events separated by an infinite time. It is shown there that a proper regularization subtracts these double-counted two-particle processes and leads to a finite result. Since we focus on the zero-temperature limit, such a regularization procedure is not necessary. 

At $p_1\gg q_0$, we can neglect all the terms in the transition probability $w_q$ involving interactions with momentum transfer $q$ of the order of $p_1$. 
For small angles $\phi < \pf/p_1$ and $\phi^{\prime}<q_0/p_1$, the square of the matrix element can then  be approximated by
\begin{equation}
w_q(\phi,\phi^{\prime}) \simeq 2m^2V_0^2\left(V_0^{\prime\prime}\right)^2\left[\phi^4+14\phi^2\left(\phi^{\prime}\right)^2+\left(\phi^{\prime}\right)^4\right].
\label{Eq:Msquared1Dspin}
\end{equation}
The divergence at $\phi=\phi^{\prime}$ does not manifest itself here because the singular terms contain the interaction potential at a large momentum transfer. 
 Since the integration region over $\phi^{\prime}$ does not contain the singular points (see above) and the potential is assumed to decay sufficiently fast, such terms would only give a  small correction.  It is thus indeed legitimate to drop them in Eq.~\eqref{Eq:Msquared1Dspin}.
   For values of $\phi^{\prime}$ much larger than $q_0/p_1$ (modulo $2\pi/3$), the squared matrix element is strongly suppressed. Comparing Eq. (\ref{Eq:Msquared1Dspin}) to Eq. (\ref{Eq:MSpinlessScalingSmall}), we thus see that the presence of spin merely  enhances the transition probability by a factor of $(\phi^\prime/\phi)^2\sim (q_0/\pf)^2$ without affecting its scaling with momentum $p_1$.
Accordingly, for the relaxation rate we find the same  scaling with  $p_1$ as in the spinless case,
\begin{equation}
\frac{1}{\tau_{p_1}}\sim  \frac{m^3  \pf^2q_0 V_0^4}{p_1^5}, \qquad p_1\gg q_0 \gg \pf \quad \text{(spinful)}.
\label{Eq:1D-high-energy-spinful}
\end{equation}

For intermediate momenta $p_1$ smaller than the characteristic scale of the interaction $q_0$ but much larger than the Fermi momentum $\pf$, all momentum transfers are small. The interaction potential $V(q)$ can thus be expanded to second order. For the square of the matrix element, we obtain
\begin{equation}
w_q(\phi,\phi^{\prime})\sim\frac{m^2 V_0^4}{q_0^4}f(\phi,\phi^{\prime}),
\label{Eq:1D-squared-matrix-element-large_q0-spinful}
\end{equation} 
where $f$ is a function of the angles $\phi$ and $\phi^{\prime}$ but is independent of $q$. This function has second order poles at $\phi=\phi^{\prime} \mod 2\pi/3$. However, at zero temperature, those poles are excluded from the integration region of $\phi^{\prime}$ by the Fermi functions of the final states [see also discussion below Eq. \eqref{Eq:1Dspinless_low_p1}]. The typical width of the excluded region is of the order of $\pf/p_1$. This leads to the scaling
\begin{equation}
\frac{1}{\tau_{p_1}}\sim \frac{m^3\pf V_0^4}{q_0^4}p_1, \qquad \pf \ll p_1 \ll q_0 \quad \text{(spinful)}.
\label{Eq:1D-intermediate-spinful}
\end{equation}
The mismatch between the scalings \eqref{Eq:1D-high-energy-spinful} and \eqref{Eq:1D-intermediate-spinful} at $p_1=q_0$ is due the neglect of terms containing large momentum transfers and poles in $\phi^{\prime}$ in the derivation of \eqref{Eq:1D-high-energy-spinful} [see discussion below Eq.~\eqref{Eq:Msquared1Dspin}]. This means that there is in fact an additional intermediate regime whose boundaries are however non-universal since they depend on the large-$q$ behavior of $V$.

In the limit of low energies we obtain for fermions with spin
\begin{equation}
\frac{1}{\tau_{p_1}}\sim\frac{m^3V_0^4}{q_0^4}(p_1-\pf)^2, \quad p_1-\pf \ll \pf \ll q_0 \quad (\mathrm{spinful}).
\label{Eq:1D-low-energy-spinful}
\end{equation}
Details of the derivation are presented in Appendix \ref{App:1D-low-energy}. Apart from logarithmic factors, this result agrees with the one obtained in Ref.~\onlinecite{KarzigGlazmanOppen10}  for the case of unscreened Coulomb interaction. Interestingly, the scaling with energy is the same as for the usual Fermi liquid in 3D. However, for a weak interaction $V_0$, the prefactor is much smaller than in the Fermi-liquid result, since the leading contribution in 1D comes only from three-particle processes.

\begin{figure}
\centering
\includegraphics[width=0.95\columnwidth]{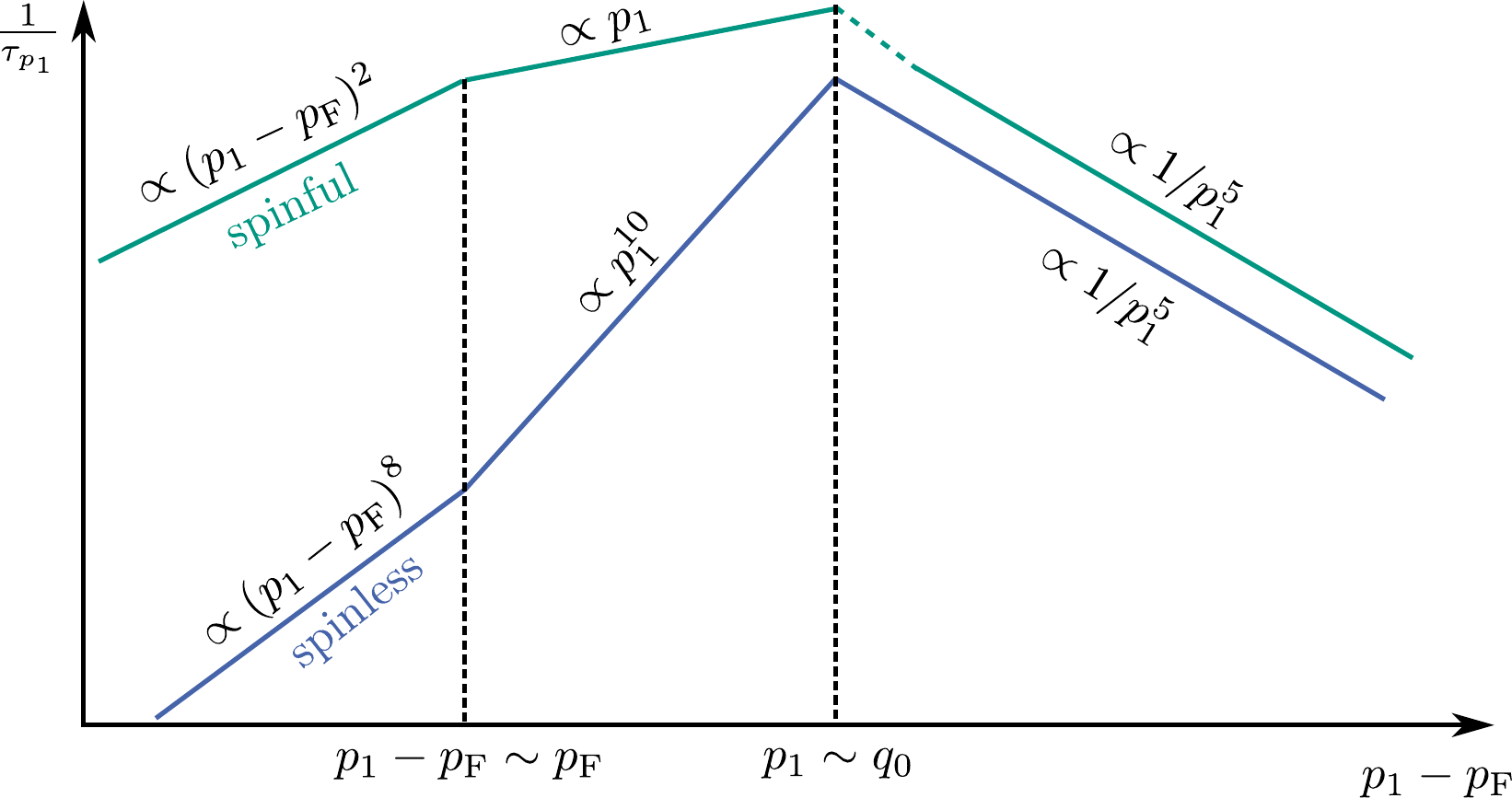}
\caption{Schematic behavior of the relaxation rate in 1D on the log-log scale for spinful and spinless fermions. At low energies, the relaxation rate is given by Eq.~\eqref{Eq:1D-low-energy-spinful} for spinful fermions and by Eq.~\eqref{Eq:1Dspinless_low_p1} for spin-polarized fermions, with a strong suppression in the latter case due to the Hartree-Fock cancellation. In the intermediate regime, $\pf\ll p_1 \ll q_0$, the relaxation rate grows as $p_1$ [see Eq.~\eqref{Eq:1D-intermediate-spinful}] for spinful and as $p_1^{10}$ [see Eq.~\eqref{Eq:1Dspinless_large_q0}] for spinless fermions. At high energies, $p_1\gg q_0$, the relaxation rate decays as $1/p_1^5$ in both cases, with a prefactor for spinless fermions that is smaller by the factor $\pf^2/q_0^2$ as compared to that for fermions with spin, see Eqs.~\eqref{Eq:1DspinlessFinal} and \eqref{Eq:1D-high-energy-spinful}, respectively. In the spinful case there is a narrow non-universal crossover regime near $p_1 \sim q_0$ marked in the plot by a dashed line.}
\label{Fig:schematic_1D_relaxation_rate}
\end{figure}

The energy dependence of the relaxation rate in 1D is schematically depicted  in Fig.~\ref{Fig:schematic_1D_relaxation_rate} for both cases of spinless and spinful fermions. 
We see that the non-monotonic behavior of the relaxation rate and the revival of the coherence at high energies observed in higher dimensions persists also in the 1D case.

\begin{figure}
\centering
\includegraphics[width=0.95\columnwidth]{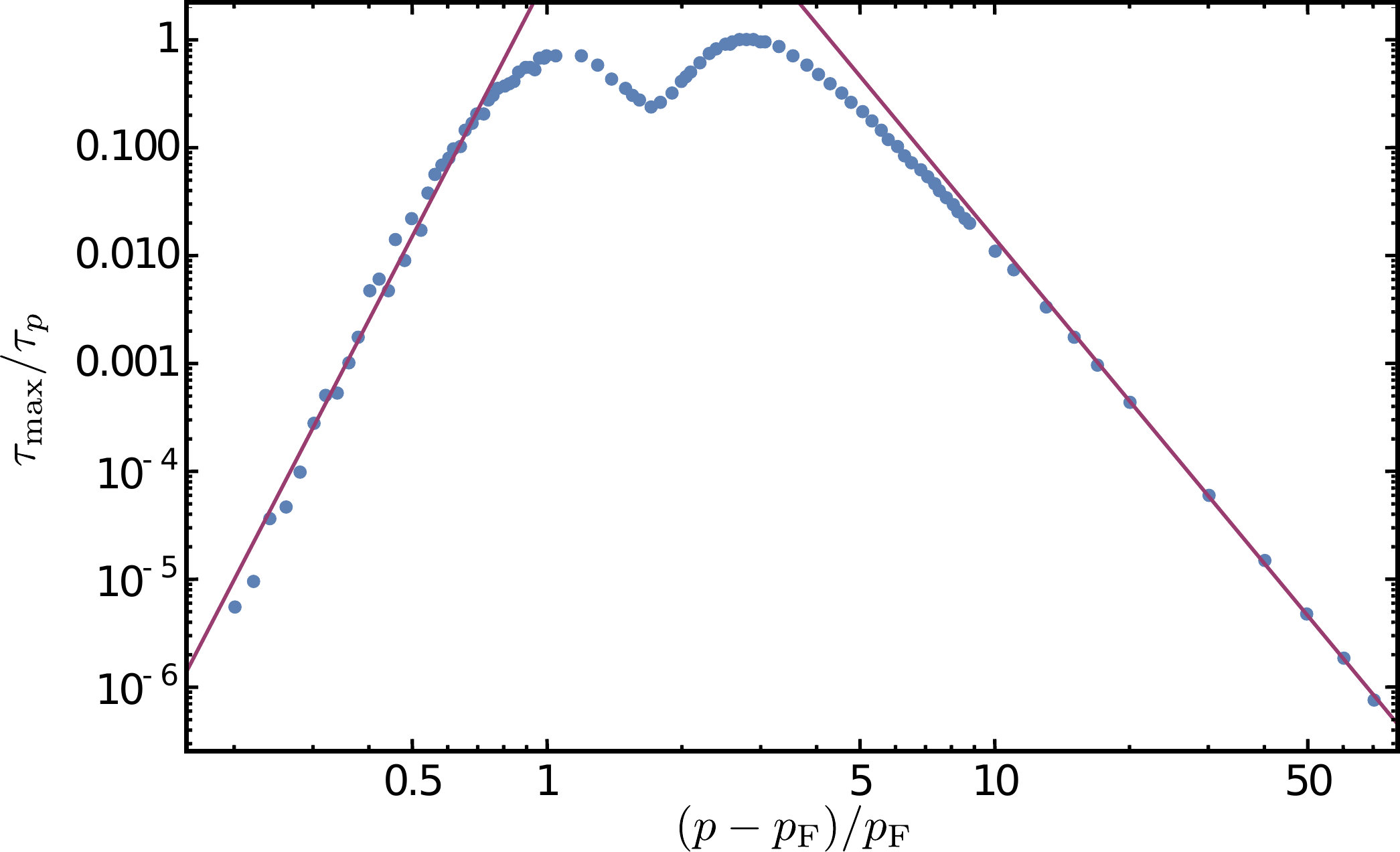}
\caption{Log-log plot of the relaxation rate $1/\tau_p$ for 1D spinless fermions  as given by numerical evaluation of Eq.~\eqref{Eq:relaxation-rate-1D-convenient} with a model interaction $V(q)= \exp(-q^2/\pf^2)$ corresponding to $q_0 = \pf$. A small temperature was used, $T/\epsilon_F=0.01$, to smear the Fermi-function steps. 
Straight lines correspond to the small-momentum and high-momentum asymptotics,  Eqs.~\eqref{Eq:1Dspinless_low_p1}  and   \eqref{Eq:1DspinlessFinal}.
}
\label{Fig:1D_relaxation_rate}
\end{figure}

In Fig. \ref{Fig:1D_relaxation_rate}, we show the results of a numerical evaluation of the relaxation rate $1/\tau_p$ of 1D spinless fermions as given by Eq.~\eqref{Eq:relaxation-rate-1D-convenient}. We have used a model interaction $V(q)= \exp(-q^2/\pf^2)$ with a characteristic scale $q_0$ equal to $\pf$. 
In this case the above analysis predicts the $(p-\pf)^8$ scaling of the relaxation rate for $p-\pf \ll \pf$ and $p^{-5}$ scaling for $p \gg \pf$
[Eqs.~\eqref{Eq:1Dspinless_low_p1}  and   \eqref{Eq:1DspinlessFinal}, respectively].  The numerical results are in very good agreement with these analytical predictions.  Interestingly, the rate $1/\tau_p$ shows also a local minimum in the crossover regime (at $p \simeq 3\pf$).

Before closing this section, let us discuss the generalizations of our results to the case of Coulomb interaction. Interestingly  the pure 1D Coulomb interaction $e^2 \ln 1/a|q|$ nullifies identically\cite{ristivojevic13} the three-particle matrix element (\ref{Eq:MTrippleFinal}) although spinless fermions with Coulomb interaction do not belong to the list of integrable models. Therefore, in our model we take into account the short-distance cutoff for 1D Coulomb interaction $d\equiv 1/q_0$ (finite width  of the 1D channel) and approximate the Coulomb interaction in momentum space by 
\begin{equation}
V_{\rm 1D}(q)=e^2\int \diff x\, e^{i qx}\frac{1}{\sqrt{x^2+1/q_0^2}}= 2e^2 K_0({|q|}/q_0)
\label{Eq:V1D}
\end{equation}
where $K_0$ is the modified Bessel function.  

At $q\gg q_0$, the interaction (\ref{Eq:V1D}) decays exponentially and in this sense does not differ from a generic model interaction $V(q)$ studied in the rest of this section. In particular, for spinless fermions, the scaling (\ref{Eq:MSpinlessScaling}) of the three-particle matrix element remains valid also in this case up to the replacement of the diverging constant $V(0)$ by 
\begin{equation}
V\left[\frac{q_0^2}{2q}\left(\tilde{\phi}^2-(\tilde{\phi}^\prime)^2\right)\right]\simeq -2e^2\ln\frac{q_0\left|\tilde{\phi}^2-(\tilde{\phi}^{\prime})^2\right|}{q}.
\end{equation}
Focusing now  on the regime $\tilde{\phi}, \tilde{\phi}^\prime\lesssim 1$, we find
\begin{equation}
M_{q}\left(q_0\tilde{\phi}/q,  q_0\tilde{\phi}^\prime /q \right)\sim \frac{m e^4}{q^2}\ln \frac{q}{q_0}
\ln \left|\frac{\tilde{\phi}-\tilde{\phi}^\prime}{\tilde{\phi}+\tilde{\phi}^\prime}\right|, \quad q\gg q_0.
\label{Eq:MCoulombLarge}
\end{equation}
Substituting this estimate for the matrix element in Eq. (\ref{Eq:rate1DSemiFinal}), we obtain the following result for the relaxation rate at highest momenta:
\begin{equation}
\frac{1}{\tau_{p_1}}\sim \frac{m^3 \pf^3 e^8\ln^2 p_1/q_0}{ p_1^5}, \qquad p_1\gg q_0\gtrsim \pf \quad \text{(spinless)}.
\label{Eq:1DspinlessCoulombFinal}
\end{equation}
We thus see that the singular nature of the Coulomb interaction potential leads in this regime to  the enhancement of relaxation rate compared to the regular case, Eq. (\ref{Eq:1DspinlessFinal}), by a factor $(q_0/\pf)\ln^2p_1/q_0$ but does not alter its $1/p_1^{5}$ scaling . 

At lower momenta of the incident particle, $p_1\lesssim q_0$,  one can approximate the Coulomb potential (\ref{Eq:V1D}) by
\begin{equation}
V_{\rm 1D}(q)\simeq e^2\left(1+\frac{q^2}{4q_0^2}\right)\ln \frac{q_0}{|q|}.
\end{equation}
The leading behavior of the three-particle matrix element, Eq. (\ref{Eq:MTrippleFinal}), then reads \cite{ristivojevic13}
\begin{eqnarray}
M_q(\phi, \phi^\prime) &\sim& 
\frac{m e^4 \ln q/q_0}{q_0^2}\sum_{k=1}^{3}\frac{\ln |\sin\frac{\phi-\phi^\prime+2\pi k/3}{2}|}{1+2\cos \left(\phi-\phi^\prime+2\pi k/3\right)} \nonumber \\ 
&-&
\left \langle \phi^\prime\rightarrow-\phi^\prime\right \rangle.
\label{Eq:MCoulobMatveev}
\end{eqnarray}
It turns out that the dominant contribution to the relaxation rate at $p_1\ll q_0$ still comes from the region of small angles $\phi^\prime$ in the integral (\ref{Eq:rate1DSemiFinal}) where Eq. (\ref{Eq:MCoulobMatveev}) reduces to [cf. Eq. (\ref{Eq:MCoulombLarge})]
\begin{equation}
M_{q}\left(\phi,  \phi^\prime \right)\sim \frac{m e^4}{q_0^2}\ln \frac{q}{q_0}
\ln \left|\frac{\phi-\phi^\prime}{\phi+\phi^\prime}\right|, \quad q\ll q_0.
\label{Eq:MCoulombLarge2}
\end{equation}
This yields the following behavior of the relaxation rate
\begin{equation}
\frac{1}{\tau_{p_1}}\sim \frac{m^3 \pf^3 e^8\ln^2 p_1/q_0}{ q_0^4 p_1}, \quad  \pf\ll p_1 \ll q_0 \quad \text{(spinless)}.
\label{Eq:1DspinlessCoulombFinal2}
\end{equation}

Finally, at low energies we can use the result of Ref.\onlinecite{ristivojevic13}, which reads (in our notations):
\begin{equation}
\begin{split}
\frac{1}{\tau_{p_1}}\sim \frac{m^3e^8\ln^2\pf/q_0}{q_0^4\pf^2}(p_1-\pf)^4, &\quad p_1-\pf \ll \pf \ll q_0.
\\
&\hspace{0.3cm}\mathrm{(spinless)}.
\end{split}
\label{Eq:1DspinlessCoulombFinal3}
\end{equation}
One can observe a parametric enhancement in  comparison with the corresponding result for a short-ranged interaction, Eq.~\eqref{Eq:1Dspinless_low_p1}.

Let us now turn to the analysis of the spin-unpolarized fermions with Coulomb interaction.
In full analogy with the case of a short-range interaction potential, in the regime $p_1\gg q_0\gg \pf$ one can focus on the region $\phi, \phi^\prime\lesssim q_0/q$  and neglect all the terms in the transition probability $w_q(\phi, \phi^\prime)$ that contain the interaction potential at a momentum transfer $q\sim p_1$. The leading behavior of the transition probability is then given by
\begin{widetext}
\begin{multline}
w_{q}\left(q_0\tilde{\phi}/q,  q_0\tilde{\phi}^\prime /q \right) \sim \frac{m^2e^8\ln^2 q/q_0}{q^4}
\left\{\left[V_{\rm 1D}(q_+)-q_+V^\prime_{\rm 1D}(q_+)-V_{\rm 1D}(q_-)+q_-V^\prime_{\rm 1D}(q_-)\right]^2\right.\\+\left[V_{\rm 1D}(q_+)-q_+V^\prime_{\rm 1D}(q_+)\right]\left[V_{\rm 1D}(q_-)-q_-V^\prime_{\rm 1D}(q_-)\right]\Big\}.
\label{Eq:wSpinFulCoulombSmall}
\end{multline}
\end{widetext}
Here $V_{1D}$ is given by Eq. (\ref{Eq:V1D}) and the momenta $q_\pm={\sqrt{3}}q_0(\tilde{\phi}\pm\tilde{\phi}^\prime)/{2}$.  As in the case of short-range interaction potential, the transition probability 
(\ref{Eq:wSpinFulCoulombSmall}) does not have any poles. Setting $\tilde{\phi}\sim \pf/q_0\ll 1$ in Eq. (\ref{Eq:wSpinFulCoulombSmall}) and using Eq. (\ref{Eq:rate1DSemiFinal}), we find
\begin{equation}
\frac{1}{\tau_{p_1}}\sim \frac{m^3 \pf^2 q_0 e^8\ln^2 p_1/q_0}{ p_1^5}, \qquad p_1\gg q_0\gtrsim \pf \quad \text{(spinful)}.
\label{Eq:1DspinfulCoulombFinal}
\end{equation}

At smaller values of $p_1$, $\pf\ll p_1\ll q_0$, we can approximate the Coulomb interaction  (\ref{Eq:V1D}) by
\begin{equation}
V_{\rm 1D}(q)\simeq e^2 \ln \frac{q_0}{|q|},
\label{Eq:unscreened_Coulomb_1D}
\end{equation}
which results in 
\begin{widetext}
\begin{equation}
w_q(\phi, \phi^\prime)\sim \frac{m^2 e^8 \ln^2q/q_0}{q^4(\cos 3\phi-\cos3\phi^\prime)^2}\left[\sum_{j=0}^2\sin\left(\frac{\phi+\phi^\prime}{2}+\frac{2\pi j}{3}\right)\ln \left|\sin\left(\frac{\phi+\phi^\prime}{2}+\frac{2\pi j}{3}\right)\right|\right]^2+\left\langle \phi^\prime\rightarrow -\phi^\prime\right\rangle.
\label{wq-coulomb-spin}
\end{equation}
\end{widetext}
Here we retain the leading behavior of the transition probability in the large factor $\ln q_0/q$. 
At small $\phi$ and $\phi^\prime$, Eq.~\eqref{wq-coulomb-spin} reduces to 
\begin{equation}
w_q(\phi, \phi^\prime)\sim  \frac{m^2 e^8 \log^2q/q_0}{q^4}\left[\frac{\ln^2 |\phi-\phi^\prime|}{(\phi+\phi^\prime)^2}+\left\langle\phi^\prime\rightarrow -\phi^\prime\right\rangle\right].
\label{Eq:wCoulombSpinfullSmall}
\end{equation}
 Substituting Eq. (\ref{Eq:wCoulombSpinfullSmall}) into Eq. (\ref{Eq:rate1DSemiFinal}) and cutting off the divergence near $\phi^\prime =\phi$ at $\phi^\prime-\phi\sim \pf/ p_1$, we get
\begin{equation}
\frac{1}{\tau_{p_1}}\sim \frac{m^3e^8 \pf}{p_1^3}\ln^2 \frac{p_1}{q_0} \ln^2 \frac{\pf}{p_1}, \quad 
\pf\ll p_1\ll q_0,\quad (\mathrm{spinful}).
\label{Eq:1DspinfulCoulombFinal2}
\end{equation}

The low-energy regime in the case of unscreened Coulomb interaction [Eq.~\eqref{Eq:unscreened_Coulomb_1D}] for spinful fermions is discussed in Ref.~\onlinecite{KarzigGlazmanOppen10}. The result reads (in our notations):
\begin{equation}
\begin{split}
\frac{1}{\tau_{p_1}}\sim &\frac{m^3e^8\ln^2\frac{q_0}{\pf}\ln^2\frac{p_1-\pf}{\pf}}{\pf^4}(p_1-\pf)^2, 
\\
&\quad p_1-\pf \ll \pf \ll q_0 \quad \text{(spinful)}.
\end{split}
\label{Eq:1DspinfulCoulombFinal3}
\end{equation} 
The leading behavior $\sim (p_1-\pf)^2$ is the same as for a short-range interaction potential [cf. Eq.~\eqref{Eq:1D-low-energy-spinful}]. This dependence results from a phase-space contribution $(p_1-\pf)^4$ and a strongly enhanced squared matrix element that contributes a factor $(p_1-\pf)^{-2}$ (see  Appendix~\ref{App:1D-low-energy}).

As is seen from Eqs.~(\ref{Eq:1DspinlessCoulombFinal}), (\ref{Eq:1DspinlessCoulombFinal2}), (\ref{Eq:1DspinlessCoulombFinal3}), (\ref{Eq:1DspinfulCoulombFinal}), (\ref{Eq:1DspinfulCoulombFinal2}) and (\ref{Eq:1DspinfulCoulombFinal3}), the behavior of the relaxation rate is strongly non-monotonic also for the case of Coulomb interaction (\ref{Eq:V1D}) in 1D systems. Specifically, for the spin-polarized electrons, $1/\tau$ increases as $(p-\pf)^4$ for low momenta but then decays as $1/p$ and eventually as  $1/p^5$ at high momenta. In the spinful case the initial increase of relaxation rate   $1/\tau\propto (p-\pf)^2$ is followed up by $1/p^3$ and, eventually, by $1/p^5$ decay. Clearly, in a physical wire, the purely 1D analysis is valid at high momenta only as long as the energy of the hot particle is below the bottom of the second band of transverse quantization.

Just as in the case of higher dimensions (see a comment in the end of Sec.~\ref{Sec:2D-3D}), the precise $1/p^5$ form of the decay of $1/\tau(p)$ relies on the parabolicity of the spectrum. However, on the qualitative level, the decay of the relaxation rate with $p$ is a much more generic feature and remains applicable as long as the velocity is increasing with momentum.


\section{Summary and discussion}
\label{Sec:Summary}

To summarize, we have studied the energy dependence of the relaxation rate of hot particles in an interacting fermionic system. We have shown that, quite generally, the relaxation rate $1/\tau $ decays according to a power law with increasing energy $\epsilon$ of a fermion in the high-energy regime. In combination with the increase of the relaxation rate with $\epsilon$  at low energies, this implies a non-monotonic dependence $1/\tau(\epsilon)$. 
 In other words, ultra-hot electrons (whose energies are much higher than the Fermi energy) recover their coherence with increasing energy. 
 
 More specifically, we have found that, for systems of spatial dimensionality $D \ge 2$, the relaxation rate $1/\tau$ scales as $p^{-1}$ at high momenta $p$, see
Eq.~ \eqref{Eq:relaxation-rate-2D-3D-final}.
 This result holds under the assumptions that the spectrum is parabolic and the interaction decays sufficiently fast at high momenta. Importantly, the Coulomb interaction belongs to this category, so that the results apply in this case as well, see Eq.~\eqref{coulomb}. The origin of the $1/p$ scaling is related to the corresponding increase of the velocity of the hot particle in comparison with those from the Fermi sea. Because of the inability of the interaction to transfer momenta larger than a characteristic scale $q_0$, the momentum transfer turns out to be nearly perpendicular to the hot-particle momentum, with a deviation of the order of $q_0/p_1$. 
 The behavior of the relaxation rate in a 3D system in the whole range of momenta, from the Fermi-liquid behavior to the ultra-hot regime, is shown schematically 
in Fig.~\ref{Fig:3DSchematic}, both for  spinful and spinless (spin-polarized) particles.  The non-monotonic behavior of $1/\tau$  is also visualized in  Fig.~\ref{Fig:Isotropic_rate} where a numerically evaluated relaxation rate is shown.

Motivated by the experiment Ref.~\onlinecite{Beidenkopf2017}, we have further studied the hot-electron relaxation in quasi-1D systems, i.e. in wires with multiple bands of transverse quantization. We have found essentially the same results as in bulk systems, see Eqs.~\eqref{wires-2D-results}  and \eqref{wires-3D-results} for quantum wires with one and two transverse dimensions, respectively. There is, however, a non-universal regime at very high energies, for the case of an almost transversal direction of the hot-particle momentum [see Fig.~\ref{Fig:non-universal-regime}]. In this regime, the discreteness of the transversal energy spectrum is essential,  resulting in a decay of the relaxation rate that is faster than $1/p$.

We have demonstrated that the non-monotonic behavior of $1/\tau$ applies also to single-channel wires where the relaxation is controlled by triple collisions. Furthermore, in this case, the decay of relaxation rate at high momenta turns out to be particularly fast, $1/\tau \propto 1/p^5$, see Eq.~\eqref{Eq:1DspinlessFinal}. This fast decay originates from (i) a factor $1/p$ related to the velocity mismatch and the limited possible momentum transfer of the interaction and (ii) a factor $1/p^4$ resulting from partial cancellation in the three-particle matrix element. The overall behavior of $1/\tau(p)$ for spinful and spinless fermions in a 1D system is shown schematically in Fig.~\ref{Fig:schematic_1D_relaxation_rate}.  Numerical evaluation of $1/\tau$ confirms our analytical findings, see Fig.~\ref{Fig:1D_relaxation_rate}.
We have also analyzed the case of Coulomb interaction in single-channel 1D geometry and demonstrated that the decay of the relaxation rate at high energies remains applicable in this situation as well.

Our findings for the 1D geometry should be contrasted to the results of Ref.~\onlinecite{Tan2010} (see also Ref.~\onlinecite{Ristivojevic2016}) where the relaxation of high-energy quasiparticles in a 1D Bose gas was studied. It was found that the relaxation rate saturates at a finite value in the limit of high energy. We offer the following explanation for the difference between this result and our $1/p^5$ prediction for the fermionic relaxation rate.  First, the factor $1/p^4$ that resulted from  matrix element cancellations in our analysis is associated with the Fermi statistics and thus does not apply in the bosonic case. Second, the factor $1/p$ in our analysis was due to the limited momentum transfer by the interaction. On the other hand, Ref.~\onlinecite{Tan2010} assumed a contact interaction, yielding a constant relaxation rate originating from processes with energy transfer of the order of the energy of the hot particle. This explanation is supported by the fact that the second contribution to the relaxation rate identified in Ref.~\onlinecite{Tan2010}---the one originating from processes with limited energy transfer---does decay as $1/p$.

Clearly, our assumption of the parabolic spectrum does not need to hold very accurately at high energies. This may determine deviations from the predicted laws of the behavior of the relaxation rate at high momenta. 

Our results are in good agreement with the experiment of Ref.~\onlinecite{Beidenkopf2017} where the energy dependence of the coherence length $L_\phi$ of electrons in a multi-band quantum wire was measured by means of scanning tunneling microscopy (STM). It was found (see Fig.~2 of Ref.~\onlinecite{Beidenkopf2017}) that $1/L_\phi$ is $ \simeq 70$\:nm near the Fermi energy, then rapidly drops down with increasing energy, shows a minimum $L_\phi \simeq 20$\:nm around the energy $E \simeq 80$\:meV (as counted from the Fermi energy), and rapidly increases with further increase of energy. In particular, at the highest energy $E \simeq 80$\:meV shown in Fig.~2b of Ref.~\onlinecite{Beidenkopf2017}, the coherent oscillations do not show any essential decay on the length scale of $ 40$\:nm, thus implying $L_\phi \gtrsim 100$\:nm. (This data point is outside the range of the $L_\phi$ axis in Fig.~2c of Ref.~\onlinecite{Beidenkopf2017}.)

Our theory predicts a $1/p$ behavior of the relaxation rate $1/\tau$ at high momenta $p$ in multi-channel quantum wires with parabolic dispersion. Since the corresponding mean free path is $\ell = v \tau$, with the velocity $v = p/m$, we get the scaling $\ell(\epsilon) \propto \epsilon$ in this regime. Thus, our model predicts that the mean-free path of hot electrons in a multimode quantum wire, as determined by electron-electron collisions, increases linearly with energy $\epsilon$. 
Assuming that the dominant factor controlling the oscillation decay in an STM experiment as carried out in Ref.~\onlinecite{Beidenkopf2017} is the relaxation due to electron-electron interaction, we obtain the same prediction for $L_\phi$.  This prediction is in a good agreement with observations of Ref.~\onlinecite{Beidenkopf2017}. Indeed, as we have just discussed, $L_\phi$ shows there an increase from $\simeq 20$\:nm to $\simeq 100$\:nm when the energy increases from $80$\:meV to $330$\:meV.  

Our predictions of non-monotonic behavior of the relaxation rate $1/\tau$ are in agreement with theoretical considerations of Ref.~\onlinecite{Beidenkopf2017}.
It is worth pointing out, however, that our theory goes beyond  the analysis of Ref.~\onlinecite{Beidenkopf2017}. Specifically, Ref.~\onlinecite{Beidenkopf2017} focused on the quasi-1D geometry and attributed the non-monotonic behavior to the lowest transverse subband of a quantum wire.  We have explored systems of various geometry and demonstrated that the non-monotonicity of $1/\tau$ is a rather general phenomenon. We have also shown that in a quantum wire the non-monotonic behavior is valid independently of the transverse subband in which the hot electron resides, at least for the case of a parabolic spectrum. We also note that, in our analysis, the relaxation of hot electrons in a multichannel quantum wire is controlled by two-electron collisions independently of the subband, at variance with Ref.~\onlinecite{Beidenkopf2017} where the relaxation in the lowest subband was attributed to three-particle collisions. It should be emphasized, however, that while we have worked within the approximation of a parabolic spectrum, the theoretical analysis of Ref.~\onlinecite{Beidenkopf2017} aimed at incorporating more accurately the band structure of InAs nanowires. 

On a more general note, our results corroborate and extend the proposal of Ref.~\onlinecite{Beidenkopf2017} that the extended phase coherence at ultra-high energies might be utilized in various quantum-technology applications. We have shown that, while the regain of coherence is particularly strong in single-channel wires, it also applies to quasi-1D, 2D, and 3D geometries. We hope that our work will trigger experimental investigations of this effect in various setups. 

We close the paper with the following comment.
 In the present work we assumed that the system is clean, i.e. the effect of disorder on scattering can be neglected.  One may wonder how these results are modified for the case of stronger disorder. This is in fact a sufficiently complex question, and one should distinguish two cases. The first possibility is that electrons in the Fermi sea are diffusive, at least on time scales relevant for the considered inelastic scattering process. It is known that in such situation the inelastic scattering (decoherence) rate of low-energy electrons is strongly enhanced by diffusive motion \cite{AltshulerAronov}. On the other hand, for high-energy electrons the relevant time scales (set by a typical energy transfer) will be shorter, so that they can remain ballistic. We thus expect that the enhancement of scattering will be less efficient for high-energy electrons, implying that the non-monotonic dependence of the decay rate should hold also for such a disordered problem. The second possibility is that the disorder is strong enough to ensure localization, which may be the case even at a finite temperature \cite{GMP2005, BAA2006}. In this case, the inelastic scattering rate at low energies will be essentially zero. On the other hand, electrons with sufficiently high energies will have a finite relaxation rate \cite{gornyi17}. Again, one can ask how this rate depends on energy. A detailed analysis of the energy dependence of the relaxation rate in a disordered system (in either diffusive or localized regime) constitutes an interesting direction for future research.

\begin{acknowledgments}

We thank H. Beidenkopf for providing us with results of Ref.~\onlinecite{Beidenkopf2017} prior to publication and for stimulating discussions of experimental data.
We are also thankful to I. V. Gornyi, N. Kainaris, and D. G. Polyakov for useful discussions. This work was supported by the Swiss National Science Foundation (Project number 200021\_163005) and by the Deutsche Forschungsgemeinschaft (Project number MI 658/9-1).

 \end{acknowledgments}


\appendix
\section{Relaxation at ultra high energies: isotropic case}
\label{App:2D_and_3D}

In this appendix, we consider scattering of a high-energy  particle by an isotropic Fermi see in $D\geq 2$ spatial dimensions and derive Eqs. (\ref{Eq:relaxation-rate-2D-3D-convenient}), (\ref{Eq:wDef}), and (\ref{Eq:Vpm}) of  Sec.~\ref{Sec:2D-3D} of the main text.

Our starting point is the Fermi golden rule expression, Eq. (\ref{Eq:relaxation-rate-2D-3D}). Making use of the rotational symmetry we first rewrite  Eq. (\ref{Eq:relaxation-rate-2D-3D}) as
\begin{widetext}
\begin{equation}
\frac{1}{\tau_{k}}=\frac{1}{mk^{D-2}S_{D-1}}\int d\pb{1} d\pb{2} d\ppb{1}d\ppb{2}\;
\delta\left(\frac{p_1^2}{2m}-\frac{k^2}{2m}\right)
\delta\left(E_{\rm i}-E_{\rm f}\right)
\cdot\delta\left({\bf P}_{\rm i}-{\bf P}_{\rm f}\right) \nf(\epsilon_2)[1-\nf(\epsilon_{1}^{\prime})][1-\nf(\epsilon_{2}^{\prime})]
\left|M_{\pb{1},\pb{2}}^{\ppb{1},\ppb{2}}\right|^2.
\label{App:Eq:rate2D_1}
\end{equation}
It is now convenient to switch to the integration over center-of-mass momentum ${\bf P}$ and the relative momenta ${\bf q}$ and ${\bf q}^\prime$ before and after the scattering
\begin{equation}
{\bf P}=\pb{1}+\pb{2}=\ppb{1}+\ppb{2}\,,\qquad
{\bf q}=\frac{\pb{1}-\pb{2}}{2}\,,\qquad  {\bf q}^\prime=\frac{\ppb{1}-\ppb{2}}{2}.
\label{App:Eq:DefPqq}
\end{equation}
After the change of integration variables we get
\begin{equation}
\frac{1}{\tau_{k}}=\frac{1}{2k^{D-2}S_{D-1}}\int \frac{d\Pb d{\bf q} d{\bf q}^\prime}{q}\;
\delta\left(\epsilon_1-\frac{k^2}{2m}\right)
\delta\left(q-q^\prime\right)
\nf(\epsilon_2)[1-\nf(\epsilon_{1}^{\prime})][1-\nf(\epsilon_{2}^{\prime})]
\left|M_{\pb{1},\pb{2}}^{\ppb{1},\ppb{2}}\right|^2.
\label{App:Eq:rate2D_2}
\end{equation}
\end{widetext}
The momenta ${\bf p}_i$ and ${\bf p}_i^\prime$ in Eq. (\ref{App:Eq:rate2D_2}) are assumed to be expressed in terms of 
${\bf P}$, ${\bf q}$ and ${\bf q}^\prime$ according to (\ref{App:Eq:DefPqq}). In particular,
\begin{equation}
\left|M_{\pb{1},\pb{2}}^{\ppb{1},\ppb{2}}\right|^2=\left[V(|{\bf q}-{\bf q}^\prime|)-V(|{\bf q}+{\bf q}^\prime|)\right]^2
\end{equation}
for spinless fermions. Due to rotational invariance ${\bf P}$ can be fixed to point in $x$-direction with the integration over direction of $P$ providing only a factor of $S_{D-1}$. It is then convenient to parametrize ${\bf q}$ and ${\bf q}^\prime$ according to
\begin{align}
{\bf q}&=q\left(\cos\phi, {\bf n}\sin\phi\right),\\
{\bf q}^{\prime}&=q\left(\cos\phi^\prime, {\bf n}^\prime\sin\phi^\prime\right)
\end{align} 
where ${\bf n}$ and ${\bf n}^\prime$ are $(D-1)$-dimensional unit vectors (perpendicular to ${\bf P}$). For the scattering rate we now get
\begin{widetext}
\begin{equation}
\begin{split}
\frac{1}{\tau_{k}}=\frac{1}{2k^{D-2}}\int_0^\infty dP dq\; P^{D-1}q^{2D-3} \int_0^\pi d\phi d\phi^\prime\left(\sin\phi\sin\phi^\prime\right)^{D-2}
\delta\left(\epsilon_1-\frac{k^2}{2m}\right)
\nf(\epsilon_2)[1-\nf(\epsilon_{1}^{\prime})][1-\nf(\epsilon_{2}^{\prime})]\\
\times \int d{\bf n} d{\bf n}^\prime\left|M_{\pb{1},\pb{2}}^{\ppb{1},\ppb{2}}\right|^2,
\end{split}
\label{App:Eq:rate2D_3}
\end{equation}
\end{widetext}
where in terms of the new integration variables
\begin{equation}
\int d{\bf n} d{\bf n}^\prime\left|M_{\pb{1},\pb{2}}^{\ppb{1},\ppb{2}}\right|^2=
\int d{\bf n} d{\bf n}^\prime \left(V_+-V_-\right)^2
\label{App:Eq:M2}
\end{equation}
with
\begin{equation}
V_{\pm}=V\left(q\sqrt{2(1\pm\cos\phi\cos\phi^\prime\pm{\bf n}\cdot {\bf n}^\prime\sin\phi\sin\phi^\prime)}\right).
\label{App:Eq:Vpm}
\end{equation}
Equations (\ref{App:Eq:rate2D_3}), (\ref{App:Eq:M2}) and (\ref{App:Eq:Vpm}) are readily seen to be equivalent to 
Eqs. (\ref{Eq:relaxation-rate-2D-3D-convenient}), (\ref{Eq:wDef}), and (\ref{Eq:Vpm}) of the main text.


\section{Relaxation rate in quasi-1D wires}

In this appendix, we present details of the calculation of the relaxation rate in multichannel quantum wires with one lateral dimension (Appendix \ref{App:Q1D-1-lateral}) and two lateral dimensions (Appendix \ref{App:Q1D-2-lateral}) discussed in Sec.~\ref{Sec:MultiBand} of the paper.

\subsection{Quasi-1D: one lateral dimension}
\label{App:Q1D-1-lateral}

\begin{figure}
\centering
\includegraphics[scale=0.4]{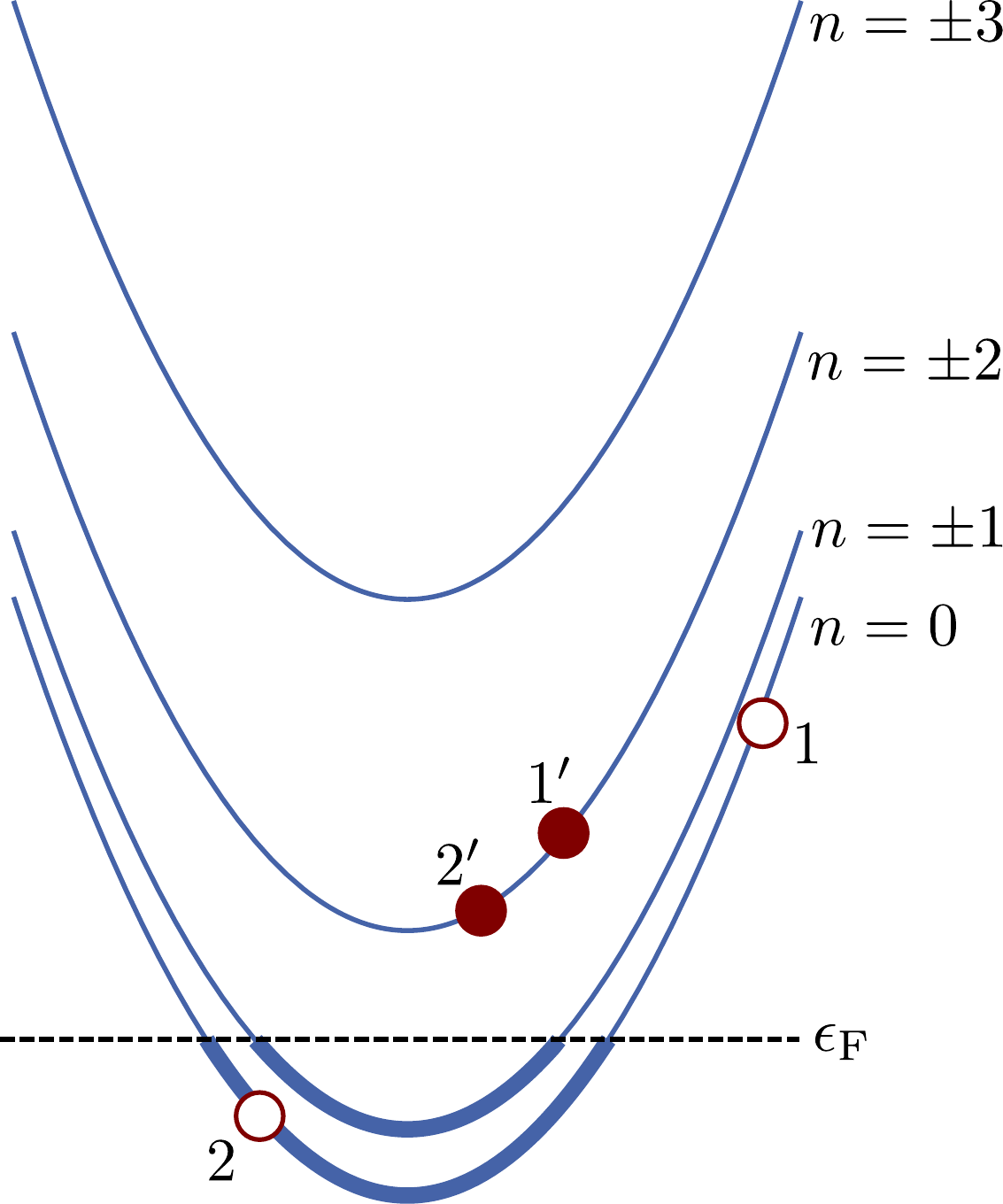}
\caption{Example of a process $12 \to 1'2'$  in a quasi-1D setup that contributes to the relaxation of the electron 1. }
\label{Fig:process-Q1D}
\end{figure}

Here we present the analysis leading to the results summarized in Eq.~\eqref{wires-2D-results} in Sec.~\ref{Sec:Q1D-1-lateral}. We start by considering a contribution of a specific process (fixed band indices $n_i$) to the relaxation rate given by Eq.~\eqref{Eq:relaxation-Q1D-1-lateral-integration-over-q}. An example of a particular process is depicted in Fig.~\ref{Fig:process-Q1D}.

Since by our assumptions there are many bands available in the energy window between the Fermi sea and the energy 
of the injected particle, the sum in Eq. (\ref{Eq:relaxation-rate-1-lateral})  is dominated by the terms where the particles $1'$ and $2'$ created in the collision process reside in otherwise empty bands. We thus have to deal with only one Fermi function requiring (at zero temperature) that  the initial cold particle $2$ is within the Fermi sea and thus limiting the range of integration over the momentum transfer $q$ to the domain defined by the inequality
\begin{equation}
\frac{1}{2m}\left(p_1-q+\frac{m\Delta_{\mathrm{eff}}}{q}\right)^2+\Delta_0 n_2^2<\ef,
\label{Eq:qRange}
\end{equation}
where $\Delta_{\rm eff}$ is defined by Eq.~\eqref{Delta-eff}.
We note that Eq. (\ref{Eq:qRange}) restricting the possible values of $q$ simultaneously limits the ``transversal energy transfer'' $\Delta_{\rm eff}$ by $|\Delta_{\mathrm{eff}}|\lesssim (p_1^2-\pf^2)/2m$.

\begin{figure}
\centering
\includegraphics[width=220pt]{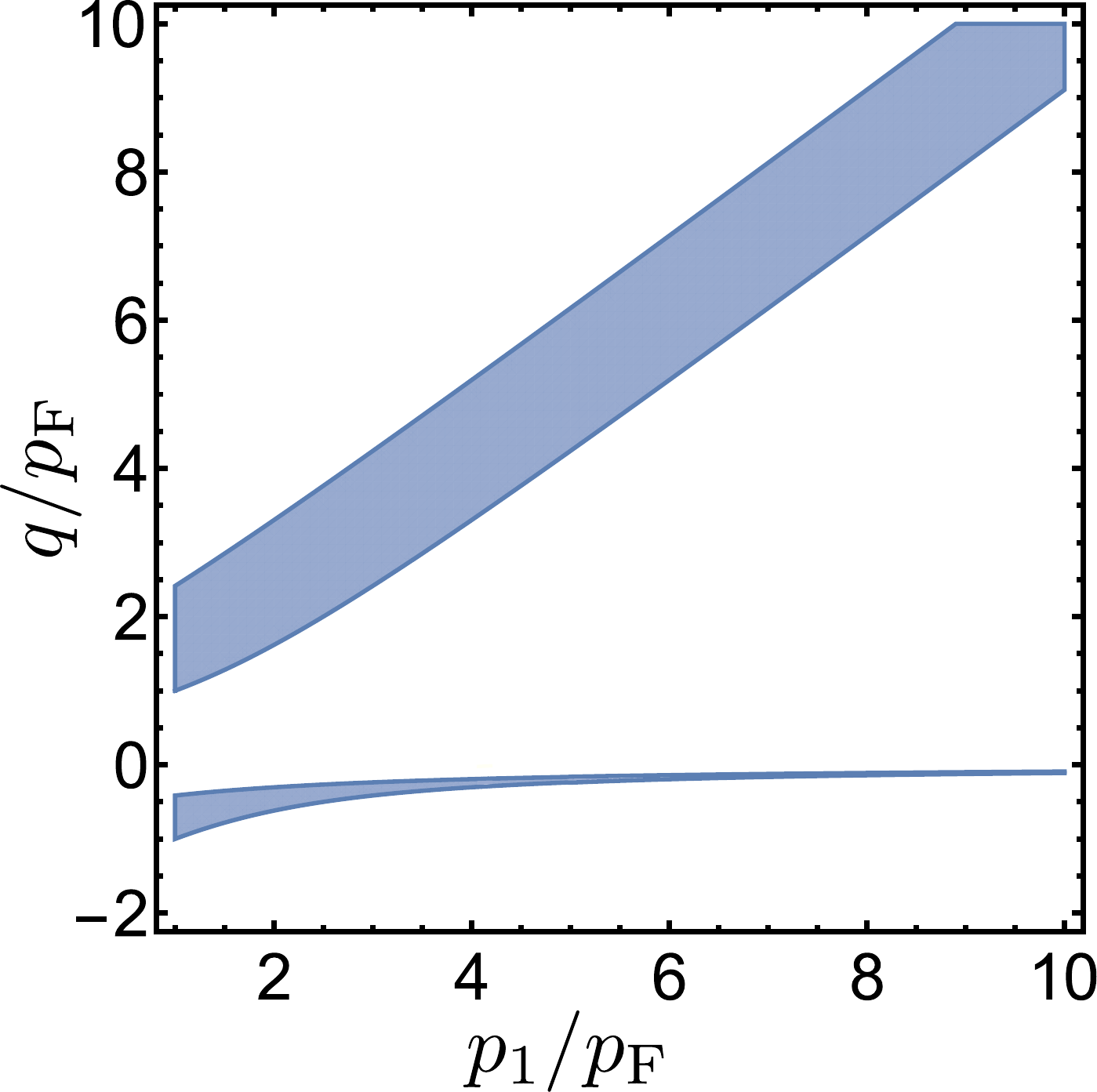}
\caption{Kinematically allowed phase space (in the plane spanned by the initial  longitudinal momentum $p_1$ and the longitudinal momentum transfer  $q=p_1-\pp{1}$) for the relaxation rate of a hot electron in a quasi-1D setup. A process with a hole created in the lowest band of transverse quantization ($n_2=0$)  and with the ``transverse energy transfer'' $\Delta_{\mathrm{eff}} = 2 \ef$ is considered.  
At large $p_1\gg\pf$, contributions come from two distinct regions: (i) $q\simeq p_1$, Eq.~\eqref{Eq:momentum-transfer-Q1D-region1}, and  (ii) small negative $q$, Eq.\eqref{Eq:momentum-transfer-Q1D-region2}. }
\label{Fig:phase-space-Q1D}
\end{figure}

In view of  the anisotropy of our model, it is not clear \emph{a priori} whether the relaxation rate of an ultra-hot electron with an energy $\epsilon \gg \epsilon_F$ depends essentially only on the energy $\epsilon$ or also on the direction of the momentum. To explore this point, we investigate two distinct limiting cases. First, we consider the case of an injected particle $1$ residing in one of the lowest bands, so that the energy $\epsilon$ is dominated by that of the longitudinal motion. In this case, the longitudinal momentum  $p_1\gg |n_1|/d$, which also implies that $p_1\gg \pf$. The opposite limit is that of a particle in one of the highly excited bands of the transversal motion, with $p_1\ll |n_1|/d$. We will show that the dependence of the relaxation rate of an ultra-hot electron on its energy is in fact almost the same in both cases. The only difference arises at very high energies, where for electrons moving in the transversal direction the discreteness of the spectrum becomes important.

Let us first assume that the relaxing particle is in one of the lowest bands and  $p_1\gg \pf$. In this case, momentum transfers $q$ close to $p_1$ and close to zero  contribute to the relaxation process, see Fig.~\ref{Fig:phase-space-Q1D}. More specifically, the allowed ranges of the momentum transfer are given by
\begin{equation}
p_1-\pft(n_2)<q<p_1+\pft(n_2)
\label{Eq:momentum-transfer-Q1D-region1}
\end{equation}
and
\begin{equation}
-\frac{m\Delta_{\mathrm{eff}}}{p_1}(1+\delta)<q<-\frac{m\Delta_{\mathrm{eff}}}{p_1}(1-\delta),
\label{Eq:momentum-transfer-Q1D-region2}
\end{equation}
where
\begin{equation}
\delta=\frac{\mathrm{sign}(\Delta_{\mathrm{eff}})}{p_1}\pft(n_2)
\end{equation}
and
\begin{equation}
\pft(n_2)=\sqrt{\pf^2-2m\Delta_0n_2^2} 
\label{Eq:pfTilde}
\end{equation}
is the (positive) momentum at which the Fermi energy intersects with the band $n_2$.

We now start with the simplest situation in which the momentum scale of the interaction is the largest momentum, $p_1 \ll q_0$. If we consider spinful fermions, we can set the matrix element constant and obtain for both branches ($q\simeq -m\Deff/p_1$ and $q\simeq p_{1}$) the scaling
\begin{equation}
\frac{1}{\tau_{\{n_i\}}}\sim m |M|^2 \frac{\sqrt{\pf^2-2m\Delta_0n_2^2}}{p_1}, \qquad \pf \ll p_1 \ll q_0.
\label{Eq:rate-Q1D-1-lateral-single-process}
\end{equation}
This is the contribution from one particular process (i.e, for given band indices). To obtain the total relaxation rate, we have to sum over all possible processes:
\begin{equation}
\frac{1}{\tau}=\sum_{n_2,n_{1^{\prime}},n_{2^{\prime}}}\frac{1}{\tau_{\{n_i\}}}\delta_{n_1+n_2,n_{1^{\prime}}+n_{2^{\prime}}}.
\label{Eq:sum_individual_processes}
\end{equation}
 The sum over $n_2$ runs over all occupied bands, $|n_2|<\sqrt{\ef/\Delta_0}=\pf d/2\pi$. It can be estimated by
\begin{equation}
\sum_{n_2=0}^{\pf d/2\pi}\sqrt{\pf^2-\frac{4\pi^2}{d^2}n_2^2}\simeq \frac{\pf^2 d}{8}. 
\label{Eq:sum-over-n2}
\end{equation}
for $\pf d\gg 1$.
The constraint related to the conservation of the transversal momentum fixes the band index $n_{2^{\prime}}$. The maximum value of the remaining band index $n_1'$ can be deduced from energy considerations: $n_{1^{\prime}}^{\mathrm{max}}\simeq n_1/2+\sqrt{n_1^2/4+p_1^2/4m\Delta_0}$. Recalling further that we assume that the longitudinal momentum of the initial hot electron $1$ dominates over the transversal one, $p_1\gg \sqrt{2m\Delta_0}n_1=2\pi n_1/d $, we obtain the relaxation rate
\begin{equation}
\frac{1}{\tau}\sim m \pf^2 d^2|M|^2\sim m V_0^2\pf^2, \qquad \pf \ll p_1 \ll q_0.
\label{Eq:scaling-1-lateral-longitudinal-direction}
\end{equation}
This rate is independent of the momentum $p_1$ of the hot electron and is identical to the result \eqref{Eq:rate-isotropic-large-q0-spin} (with $D=2$) obtained in the same range of $p_1$ for the isotropic situation in two dimensions.

For fermions without spin, the Hartree-Fock cancellation makes it necessary to analyze the specific form \eqref{Eq:interaction-Q1D} of the interaction. The momentum transfers of the direct and exchange terms are
\begin{align}
&q_{\mathrm{dir}}^2=q^2+\frac{4\pi^2}{d^2}(n_1-\np{1})^2,
\label{Eq:momentum-transfer-Q1D-direct}
\\
&q_{\mathrm{ex}}^2=\frac{16\pi^4}{q^2d^4}[(n_1-\np{1})(\np{1}-n_2)]^2+\frac{4\pi^2}{d^2}(\np{1}-n_2)^2,
\label{Eq:momentum-transfer-Q1D-exchange}
\end{align}
respectively. The momentum $q$ is integrated over the regions given by Eqs.~\eqref{Eq:momentum-transfer-Q1D-region1} and \eqref{Eq:momentum-transfer-Q1D-region2}. The maximum band index $\np{1}$ is given by $\np{1}^{\mathrm{max}}\approx p_1 d/2\sqrt{2}\pi$. For most processes $\np{1}\gg n_1,n_2$. Estimating the summation over all processes yields the scaling
\begin{equation}
\frac{1}{\tau_{\epsilon}}\sim m V_0^2 \pf^2\frac{p_1^4}{q_0^4}\sim m^3 V_0^2\frac{\pf^2}{q_0^4}\epsilon^2, \qquad \pf \ll p_1 \ll q_0,
\label{Eq:scaling-1-lateral-longitudinal-direction-p1-smaller-q0}
\end{equation}
which again agrees with the scaling \eqref{Eq:rate-isotropic-large-q0-spinless} in the corresponding regime for the isotropic 2D situation. Here, we introduced the energy $\epsilon$ of the particle with momentum $p_1$ in order to facilitate the comparison between the two situations in which the incident momentum of the particle $1$ is in the longitudinal and transversal direction, respectively. Both regions of the integration over $q$ [Eqs.~\eqref{Eq:momentum-transfer-Q1D-region1} and \eqref{Eq:momentum-transfer-Q1D-region2}] yield contributions to the scattering rate of the order of Eq.~(\ref{Eq:scaling-1-lateral-longitudinal-direction-p1-smaller-q0}). We briefly discuss the origin of the individual factors using the example of the region given by Eq.~\eqref{Eq:momentum-transfer-Q1D-region1}. We estimate the integral in Eq.~\eqref{Eq:relaxation-Q1D-1-lateral-integration-over-q} by setting $q=p_1$ and multiplying by the width $\pft(n_2)$ of the integration domain. We can neglect the $n_2$-dependence of the matrix element for most processes. According to Eq.~\eqref{Eq:sum-over-n2}, the summation over $n_2$ in \eqref{Eq:sum_individual_processes} yields a factor $\pf^2 d$. The summation of the squared matrix element over $\np{1}$ leads to the factor $(V_0^2/q_0^4d^2)\cdot p_1^5 d$. Combining these results, we get the scaling \eqref{Eq:scaling-1-lateral-longitudinal-direction-p1-smaller-q0}. 

When the longitudinal momentum exceeds the momentum scale of the interaction, $p_1\gg q_0 \gtrsim \pf$, 
one of the terms in the square brackets of Eq.~\eqref{matrix-element-no-spin} is much larger than the other. In this case the Hartree-Fock cancellation is inefficient, implying that the result is essentially the same for models with and without spin. 
Further, we find that only band indices $\np{1}$ up to $\np{1}^{\mathrm{max}}\approx q_0 d/2\pi$ contribute. For higher $\np{1}$, the transversal momentum transfer is larger than $q_0$, so that their contributions are suppressed.
The resulting scaling of the inverse lifetime is given by
\begin{equation}
\frac{1}{\tau_{\epsilon}}\sim m V_0^2 \pf^2 \frac{q_0}{p_1}\sim \sqrt{m}V_0^2\pf^2\frac{q_0}{\sqrt{\epsilon}}, \qquad \pf\lesssim q_0 \ll p_1,
\label{Eq:scaling-1-lateral-longitudinal-direction-p1-larger-q0}
\end{equation}
which is the same behavior as in the corresponding regime for the isotropic 2D scaling, Eq.~\eqref{Eq:relaxation-rate-2D-3D-final}. 
The result \eqref{Eq:scaling-1-lateral-longitudinal-direction-p1-larger-q0} is obtained from \eqref{Eq:rate-Q1D-1-lateral-single-process} and \eqref{Eq:sum_individual_processes}. The sum over $n_2$ contributes a factor $\pf^2 d$ and the summation of the squared matrix element over $\np{1}$ yields a factor $(V_0^2/d^2) \cdot q_0 d$. 

Now we turn to the situation when the transversal momentum of the initial hot electron dominates over the longitudinal one. This is the case when the hot electron resides close to the bottom of a high band: $p_1 \simeq 0$, $ \Delta _0 n_1^2 \gg \ef$.  To analyze this limit, we use the general result Eq.~\eqref{Eq:relaxation-Q1D-1-lateral-integration-over-q} and set there $p_1=0$. 
As has been pointed above, the dominant contribution originates from processes in which  the final particles $1'$ and $2'$ reside in otherwise empty bands, $n_1', n_2' > n_2^{\rm max}$. In this case, there is only one Fermi function restricting the initial cold electron to be a part of the Fermi sea. At $T=0$, this function restricts the integration over $q$  to the following two regions: 
\begin{equation}
\begin{split}
-\frac{\pft}{2}-\sqrt{\frac{\pft^2}{4}+m\Deff}<&q<\frac{\pft}{2}-\sqrt{\frac{\pft^2}{4}+m\Deff},
\\
-\frac{\pft}{2}+\sqrt{\frac{\pft^2}{4}+m\Deff}<&q<\frac{\pft}{2}+\sqrt{\frac{\pft^2}{4}+m\Deff},
\end{split}
\label{Eq:integration-region-1-lateral-transversal-direction}
\end{equation}
where $\tilde{p}_{\mathrm{F}}$ stands for $\tilde{p}_{\mathrm{F}}(n_2)$ as defined by Eq.~\eqref{Eq:pfTilde}.
For dominant processes for a relaxation of a hot particle we have $m\Deff \gtrsim \pft^2$, so that the two integration regions \eqref{Eq:integration-region-1-lateral-transversal-direction} can be approximated by 
$$-\sqrt{m\Deff}-\pft(n_2)/2<q<-\sqrt{m\Deff}+\pft(n_2)/2$$ and $$\sqrt{m\Deff}-\pft(n_2)/2<q<\sqrt{m\Deff}+\pft(n_2)/2.$$

We start again by considering the case when the momentum scale $q_0$ is large, $\pf \ll 2\pi n_1/d \ll q_0$. In the presence of spin, the matrix element can be replaced by a constant. A process with given values of band indices gives the contribution
\begin{equation}
\begin{split}
\frac{1}{\tau_{\{n_i\}}}& \simeq |M|^2\pft \sqrt{\frac{m}{\Deff}}
\\
&=|M|^2\pft \sqrt{\frac{m}{2\Delta_0(n_1-\np{1})(\np{1}-n_2)}}
\end{split}
\label{Eq:scaling-Q1D-one-process-lateral-direction}
\end{equation}
to the total relaxation rate. 
Since $n_1, n_1' \gg n_2$ for dominant processes, we can neglect the $n_2$-dependence in $\Deff$.  The summation over $n_2$ leads then to a factor $\pf^2 d$ [cf. Eq.~\eqref{Eq:sum-over-n2}]. We can now sum over $\np{1}$ using the Euler-Maclaurin formula:
\begin{align}
\frac{1}{\tau}&=\pf^2 d |M|^2 \sqrt{\frac{m}{2\Delta_0}}\sum_{\np{1}=1}^{n_1-1} \frac{1}{\sqrt{\np{1}(n_1-\np{1})}} \nonumber
\\
&\simeq \pf^2 d |M|^2 \sqrt{\frac{m}{2\Delta_0}}\int_{1}^{n_1-1} \frac{\diff x}{\sqrt{x(n_1-x)}}+\mathcal{O}(n_1^{-1/2}) \nonumber
\\
&\sim m \pf^2 d^2  |M|^2\sim m V_0^2 \pf^2, \qquad \pf \ll \frac{2\pi n_1}{d} \ll q_0.
\label{Eq:scaling-1-lateral-transversal-direction}
\end{align}
The obtained relaxation rate (\eqref{Eq:scaling-1-lateral-transversal-direction}) is identical to the result \eqref{Eq:scaling-1-lateral-longitudinal-direction} for the corresponding regime in the case of the dominant longitudinal momentum of the initial hot particle. We thus see that the relaxation rate does not essentially dependent on the direction of the momentum of the hot particle in this regime. As we are going to show, this applies also to other regimes.

In the absence of spin, the above leading contribution (with the matrix element approximated by a constant) vanishes due to Hartree-Fock cancellation. We thus need to take into account the momentum dependence of the interaction matrix element at the relevant momentum transfers that are given by \eqref{Eq:momentum-transfer-Q1D-direct} and \eqref{Eq:momentum-transfer-Q1D-exchange}. In the considered case of the initial momentum smaller than $q_0$,  the momentum transfer is automatically smaller than $q_0$, so that the interaction range does not impose further restrictions on the relevant phase space. The total relaxation rate can be computed via
\begin{equation}
\frac{1}{\tau}\sim m\pf^2 d \frac{V_0^2}{q_0^4d^5}n_1^2\sum_{\np{1}=1}^{n_1-1}\frac{(n_1-2\np{1})^2}{\sqrt{\np{1}(n_1-\np{1})}}.
\end{equation}
Estimating the sum leads to
\begin{equation}
\begin{split}
\frac{1}{\tau_{\epsilon}} \sim m V_0^2 \frac{\pf^2}{q_0^4}\left(\frac{2\pi n_1}{d}\right)^4\sim m^3& V_0^2\frac{\pf^2}{q_0^4}\cdot\epsilon^2, 
\\
 &\pf \ll \frac{2\pi n_1}{d} \ll q_0,
\end{split}
\label{Eq:scaling-1-lateral-transversal-direction-n1-smaller-q0}
\end{equation}
in full consistency with Eq.~\eqref{Eq:scaling-1-lateral-longitudinal-direction-p1-smaller-q0} for the situation when the incident momentum $\mathbf{p}_1$ is along the longitudinal direction.

If the transversal momentum is larger than the momentum scale of the interaction, $2\pi n_1/d\gg q_0 \gtrsim \pf$, the matrix element suppresses some processes. Specifically, the direct term contributes if $\np{1}>n_1-q_0^2d^2/4\pi^2n_1$, while the exchange term yields an essential contribution for $\np{1}<q_0^2d^2/4\pi^2 n_1$. Accordingly, the Hartree-Fock cancellation in the spinless situation does not occur, so that we obtain the same scaling law as in the spinful case. We neglect again the $n_2$-dependence, since for most processes $\np{1} \gg n_2$. There is a constraint $q_0^2 d^2/4\pi^2n_1>1$ for the sum over $\np{1}$ to be non-zero. Introducing the transversal momentum of the inserted particle $p_{1\perp}=2\pi n_1/d$, we can rewrite this condition as $p_{1\perp}<q_0^2 d/2\pi$.
This condition ensures that the typical momentum transfer in the transversal direction $q_0^2 / p_{1\perp}$ (see discussion in the end of Sec.~\ref{Sec:2D-3D}) exceeds the quantization step $2\pi/d$. 
 If this condition is not fulfilled, the relaxation rate is strongly suppressed since only the large-momentum tail of $V(q)$ contributes. Therefore, in the regime $p_{1\perp}>q_0^2 d$, the effect of the discreteness of the spectrum leads to an additional suppression for a quasi-1D system in comparison with the isotropic 2D situation. The precise form of this suppression is non-universal, and we do not discuss it here. 

Assuming that the condition $p_{1\perp}<q_0^2d$ is fulfilled, we estimate the relaxation rate. The summation over $n_2$ yields a factor $\sim \pf^2 d$, see Eq.~\eqref{Eq:sum-over-n2}. The direct and exchange terms yield the same contributions. It is thus sufficient to estimate the exchange term, which reads
\begin{equation}
\frac{1}{\tau}\sim m \pf^2 d \frac{V_0^2}{d^2}\sum_{\np{1}=1}^{\frac{q_0^2 d}{2\pi p_{1\perp}}}\frac{d}{\sqrt{\np{1}(n_1-\np{1})}}.
\end{equation}
Replacing the sum by an integral, we get
\begin{equation}
\begin{split}
\frac{1}{\tau_{\epsilon}}\sim m V_0^2 \pf^2 &\frac{q_0}{p_{1\perp}}\sim \sqrt{m} V_0^2 \pf^2 \frac{q_0}{\sqrt{\epsilon}}, 
\\
&\qquad \pf \lesssim q_0 \ll p_{1\perp} \ll q_0^2 d,
\end{split}
\label{Eq:scaling-1-lateral-transversal-direction-n1-larger-q0}
\end{equation}
in consistency with the scaling \eqref{Eq:scaling-1-lateral-longitudinal-direction-p1-larger-q0} of the relaxation rate of electrons that move in the longitudinal direction.

As explained above, in the case when the momentum of the hot electron points in the transversal direction, the relaxation rate is additionally suppressed at $p_{1\perp} \gg q_0^2 d$ due to discreteness of the spectrum. On the other hand, when the momentum is in the longitudinal direction, the only condition to recover the isotropic 2D result at high momenta $p_1$ is $q_0 d>1$. 
 In order to analyze the crossover between both limits, we consider the minimal momentum transfer in the situation when the hot particle is in a high band $n_1 \gg \pf^2 d^2$ but has also a finite momentum component $p_{1}$ in the longitudinal direction. The longitudinal momentum transfer $q=p_{1}-p^{\prime}_{1}$ can be estimated in this regime as
\begin{equation}
-p_{1}q+q^2=\frac{n_{1^{\prime}}(n_1-n_{1^{\prime}})}{d^2}.
\label{Eq:minimal-transfer}
\end{equation}
For $p_{1}=0$, we obtain for the minimal longitudinal momentum transfer $q_{\mathrm{min}}=\sqrt{n_1}/d=\sqrt{p_{1\perp}/d}$. The condition for the universal regime of the relaxation rate ($1/p_{1\perp}^{-1}$ decay) is thus given by $q_{\mathrm{min}}<q_0$, which reproduces the last condition in the last line of Eq.~\eqref{Eq:scaling-1-lateral-transversal-direction-n1-larger-q0} that limits $p_{1\perp}$ from above. On the other hand, if the longitudinal momentum component $p_{1}$ becomes larger than $q_0$, the condition for this regime is determined by the first term of the left-hand side of Eq.~\eqref{Eq:minimal-transfer}. We thus find the following condition of the universal regime in this situation: $p_{1}/p_{1\perp}>1/q_0d$. To summarize, a non-universal regime with a decay of the relaxation rate that is faster than $1/\sqrt{\epsilon}$ occurs if the longitudinal ($p_{1}$) and transversal ($p_{1\perp}$) components of momentum of particle $1$ satisfy
\begin{equation}
p_{1\perp}>q_0^2 d \quad \mathrm{and} \quad \frac{p_{1}}{p_{1\perp}}<\frac{1}{q_0 d}.
\end{equation} 

All results for quasi-1D wires with one lateral dimension are summarized in Eq.~\eqref{wires-2D-results}. 

\subsection{Quasi-1D: two lateral dimensions\label{App:Q1D-2-lateral}}
Here we present details of the analysis leading to the results summarized in Eq.~\eqref{wires-3D-results} in Sec.~\ref{Sec:Q1D-2-lateral}.
The calculations are to large extent analogous to those for one lateral dimension presented in Appendix \ref{App:Q1D-1-lateral} above.

We use Eq.~(\ref{Eq:relaxation-Q1D-1-lateral-integration-over-q}) and consider first the situation in which the longitudinal momentum dominates over the transversal one, $p_1 \gg 2\pi |\nb_1|/d$. We start with the case when the momentum scale of the interaction is large, $p_1<q_0$. In the spinful case, the contribution of a particular process with given band indices is given by Eq.~\eqref{Eq:rate-Q1D-1-lateral-single-process} with the replacement $n_2\to \nb_2$. The summation of $\nb_2$ over the occupied bands leads now to a factor $\sim \pf (\pf d)^2$. The summation over $\nbp{1}$ is limited by $|\nbp{1}|<p_1 d/2\sqrt{2}\pi$ in view of the restriction imposed by the energy conservation. The sum over $\nbp{1}$ thus leads to a factor $\sim(p_1 d)^2$. Altogether, we find in the presence of spin the following behavior of the relaxation rate:
\begin{equation}
\frac{1}{\tau_{\epsilon}}\sim m V_0^2 \pf^3 p_1\sim m^{3/2} V_0^2 \pf^3 \sqrt{\epsilon}, \qquad \pf\ll p_1\ll q_0.
\label{Eq:scaling-2-lateral-longitudinal-direction}
\end{equation}
In analogy with Sec.~\ref{Sec:Q1D-1-lateral}, we find a full agreement of this result with
the behavior of the relaxation rate \eqref{Eq:rate-isotropic-large-q0-spin} in the corresponding regime of a bulk 3D system.

In the spinless case and for large $q_0$, we need to analyze the precise form \eqref{Eq:interaction-Q1D} of the interaction at the relevant momentum transfers because of the Hartree-Fock cancellation. The momentum transfers of the direct and exchange term are given by Eqs. \eqref{Eq:momentum-transfer-Q1D-direct} and \eqref{Eq:momentum-transfer-Q1D-exchange} with a replacement $n_i \to \nb_i$. For dominant processes, we can neglect the  $\nb_1$ and $\nb_2$ dependencies of the matrix element. The $q$ integration over the regions \eqref{Eq:momentum-transfer-Q1D-region1} and \eqref{Eq:momentum-transfer-Q1D-region2} yields a factor $\sim m/p_1 \cdot \sqrt{\pf^2-2m\Delta_0 \nb_2^2}$. Summing over $\nb_2$ results in a factor $\sim \pf (\pf d)^2$. The summation of the squared matrix element over $\nbp{1}$ yields a factor $\sim V_0^2/(q_0^4d^4)\cdot p_1^6 d^2$. The inverse lifetime is thus given by
\begin{equation}
\frac{1}{\tau_{\epsilon}}\sim m V_0^2 \pf^3 \frac{p_1^5}{q_0^4}\sim m^{7/2} V_0^2 \frac{\pf^3}{q_0^4} \epsilon^{5/2}, \qquad \pf \ll p_1 \ll q_0,
\label{Eq:scaling-2-lateral-longitudinal-direction-p1-smaller-q0}
\end{equation} 
which agrees with the isotropic 3D situation, Eq.\eqref{Eq:rate-isotropic-large-q0-spinless}.

In the case $p_1 \gg q_0 \gtrsim \pf$, the matrix element restricts the summation over $\nbp{1}$ to the disk with radius $q_0 d/2\pi$. The result for the relaxation rate does not depend on the presence of spin:
\begin{equation}
\frac{1}{\tau_{\epsilon}}\sim m V_0^2 \pf^3 \frac{q_0^2}{p_1}\sim \sqrt{m}V_0^2 \pf^3 \frac{q_0^2}{\sqrt{\epsilon}}, \qquad \pf \lesssim q_0 \ll p_1.
\label{Eq:scaling-2-lateral-longitudinal-direction-p1-larger-q0}
\end{equation}
Here the factor $m/p_1 \cdot \pf (\pf d)^2$ stems from the integration over $q$ and the summation over $\nb_2$, while the summation of the squared matrix element over $\nbp{1}$ yields a factor $V_0^2/d^4 \cdot (q_0 d)^2$. This result is in accordance with the isotropic three dimensional scaling law, Eq.~\eqref{Eq:relaxation-rate-2D-3D-final}.

We discuss now the situation when the momentum of the hot electron points in a transversal direction, i.e., the electron resides at the bottom of a high band, $\Delta_0 \nb_1^2 \gg \ef$. We start again with the spinful case. Assuming first that the  transversal momenta satisfy $\pf \ll 2\pi |\nb_1|/d \ll q_0$, we use Eq.~\eqref{Eq:scaling-Q1D-one-process-lateral-direction} with a replacement $n_i \to \nb_i$ for the contribution of a process with given band indices. As before, we focus on dominant processes characterized by $\Deff>0$. 

It is worth mentioning a subtlety in the case of a wire with two transverse direction that did not exist for one transverse direction. Specifically, besides the trivial zeros $\nbp{1}=\nb_1$ and $\nbp{1}=\nb_2$ of $\Deff$, there are additional zeros that occur whenever the vector $\nb_1-\nbp{1}$ is orthogonal to $\nbp{1}-\nb_2$. These zeros lead to singularities of the relaxation rate. Within Eq.~(\ref{Eq:relaxation-Q1D-1-lateral-integration-over-q}), such zeros lead to logarithmic singularities. These singularities, however, get regularized, and the corresponding terms turn out to be of minor importance for our analysis, for a 
discussion of this point see Sec.~\ref{Sec:Q1D-2-lateral}. We thus drop these terms below.

The summation over $\nb_2$ leads again to a factor $\sim\pf(\pf d)^2$. Performing the summation over $\nbp{1}$, we neglect the $\nb_2$-dependence since for most processes $|\nbp{1}|\gg |\nb_2|$. Energetic considerations restrict the summation over $\nbp{1}$ to the disk with a radius $|\nb_1|$. 
The summation over $\nbp{1}$ can be approximated by the corresponding integration. We parametrize the denominator appearing in Eq.~\eqref{Eq:scaling-Q1D-one-process-lateral-direction}:
\begin{equation}
\frac{1}{\sqrt{(\nb_1-\nbp{1})\nbp{1}}}=\frac{1}{\sqrt{|\nb_1||\nbp{1}|\cos\phi-|\nbp{1}|^2}},
\end{equation}
where $\cos\phi>|\nbp{1}|/|\nb_1|$ to ensure the condition $\Deff>0$. The integration over $|\nbp{1}|$ and the angle $\phi$ leads to the factor $|\nb_1|$. 
Collecting everything, we find
\begin{equation}
\begin{split}
\frac{1}{\tau_{\epsilon}}\sim m V_0^2 \pf^3 \frac{2\pi |\nb_1|}{d}\sim & \,m^{3/2}V_0^2 \pf^3\sqrt{\epsilon}\:,
\\
& \qquad \pf \ll  \frac{2\pi |\nb_1|}{d} \ll q_0.
\end{split}
\label{Eq:scaling-2-lateral-transversal-direction}
\end{equation}
The formula \eqref{Eq:scaling-2-lateral-transversal-direction} for the momentum $\pb{1}$ pointing perpendicular to the wire axis
is in agreement with the result \eqref{Eq:scaling-2-lateral-longitudinal-direction} for the momentum $\pb{1}$ pointing in the longitudinal direction.

For the same conditions on the momentum scales but in the absence of spin, the momentum dependence of the matrix element is important in view of a partial cancellation of direct and exchange term. The momentum transfers of the direct and exchange terms for a process with given band indices are given by Eqs.~\eqref{Eq:momentum-transfer-Q1D-direct} and \eqref{Eq:momentum-transfer-Q1D-exchange}, respectively (with a replacement $n_i \to \nb_i$), and the $q$-integration goes over the two regions of Eq.~\eqref{Eq:integration-region-1-lateral-transversal-direction}. As before, we neglect the $\nb_2$-dependence everywhere except for the width of the region of $q$-integration. The summation over $\nb_2$ yields a factor $\sim\pf (\pf d)^2$. 
Since we consider the regime $2\pi |\nb_1|/d<q_0$, all momentum transfers allowed by kinematic restrictions are smaller than $q_0$. The sum over $\nbp{1}$ is constrained by $|\nbp{1}|<|\nb_1|$ as well as by $\nbp{1}(\nb_1-\nbp{1})>0$, which ensures $\Deff>0$, the condition valid for dominant processes. This sum scales as $|\nb_1|^5$. Collecting all factors, we obtain
\begin{equation}
\begin{split}
\frac{1}{\tau_{\epsilon}}\sim m V_0^2\frac{\pf^3}{q_0^4} \left(\frac{2\pi |\nb_1|}{d}\right)^5\sim &\,m^{7/2}V_0^2 \frac{\pf^3}{q_0^4}\epsilon^{5/2},  
\\
&\,\,\pf \ll \frac{2\pi |\nb_1|}{d}\ll q_0,
\end{split}
\label{Eq:scaling-2lateral-transversal-direction-n1-smaller-q0}
\end{equation}
which is in accordance with Eq.~\eqref{Eq:scaling-2-lateral-longitudinal-direction-p1-smaller-q0} for the case of the hot-particle momentum directed along the wire axis.

Finally, in the case when the hot-electron transverse momentum is the largest scale, $2\pi|\nb_1|/d \gg q_0\gtrsim \pf$, the direct term contributes if $(2\pi/d)^2(\nb_1^2-\nb_1\nbp{1})<q_0^2$ and the exchange term if $(2\pi/d)^2\nb_1\nbp{1}<q_0^2$. Both regions that constrain the summation over $\nbp{1}$ do not overlap in the considered regime. Thus, in this regime the presence or absence of spin is immaterial:  spinful and spinless fermions are characterized by the same scaling of the relaxation rate. There is an additional constraint $\nbp{1}(\nb_1-\nbp{1})>0$ selecting the most important processes ($\Deff>0$). 
Replacing the sum over $\nbp{1}$ by an integral, we get
\begin{equation}
\begin{split}
\frac{1}{\tau_{\epsilon}}\sim m V_0^2\pf^3q_0^2 &\left(\frac{2\pi |\nb_1|}{d}\right)^{-1}\sim \,\sqrt{m} V_0^2 \pf^3 \frac{q_0}{\sqrt{\epsilon}},
\\
 &\,\, \pf \lesssim q_0 \ll \frac{2\pi |\nb_1|}{d} \ll q_0^2d,
\end{split}
\label{Eq:scaling-2-laterlal-transversal-direction-n1-larger-q0}
\end{equation}
in full consistency with Eq.~\eqref{Eq:scaling-2-lateral-longitudinal-direction-p1-larger-q0} for the case when $\pb{1}$ points in the longitudinal direction. 
As in the case of one lateral dimension, the isotropic 3D result is reproduced in  Eq.~\eqref{Eq:scaling-2-laterlal-transversal-direction-n1-larger-q0} under the condition $q_0^2 d^2/|\nb_1|>1$. If this condition is not met, the relaxation rate decays even faster since only the large-momentum tail of the interaction potential contributes. This non-universal regime exists if the direction of the momentum of the hot particle is almost in transversal direction. More specifically, in full analogy with the analysis in the end of Appendix~\ref{App:Q1D-1-lateral}, this non-universal regime is determined by the following conditions:
\begin{equation}
\frac{|\nb_1|}{d}>q_0^2 d \quad \mathrm{and}\quad \frac{p_1}{|\nb_1|/d}<\frac{1}{q_0 d},
\end{equation}
where $p_1$ is the longitudinal component of the momentum of particle $1$. 

All the results for quasi-1D wires with two lateral dimensions are summarized in Eq.~\eqref{wires-3D-results}.

\section{Triple collisions in 1D wires with quadratic dispersion relation}
\label{App:1D-amplitude}
In this appendix we derive a convenient representation for the on-shell three-particle matrix element valid for particles with quadratic spectrum. 

 The general form of the three-particle matrix element for spinful fermions is given by the vacuum expectation value\cite{Lunde07}
\begin{equation}
M_{123}^{1^{\prime}2^{\prime}3^{\prime}}=\langle a_{1^{\prime}}a_{2^{\prime}}a_{3^{\prime}}|\hat{V}\frac{1}{E-\hat{H}_0+i0}\hat{V}|a^{\dagger}_1a^{\dagger}_2a^{\dagger}_3\rangle,
\end{equation}
where $\hat{H}_0$ and $\hat{V}$ are the free and interaction part of the Hamiltonian, $E$ is the total energy, and $a^{(\dagger)}_{i}$ annihilates (creates) a fermion with momentum $p_i$ and spin-projection $\sigma_i$. It can be recast in the form\cite{Lunde07}
\begin{equation}
\begin{split}
M_{123}^{1^{\prime}2^{\prime}3^{\prime}}=&\sum_{(abc) \in \mathcal{P}(123)}\hspace{-0.4cm}\mathrm{sgn}(123)\hspace{-0.4cm}\sum_{(a^{\prime}b^{\prime}c^{\prime})\in \mathcal{P}(1^{\prime}2^{\prime}3^{\prime})}\hspace{-0.4cm}\mathrm{sgn}(1^{\prime}2^{\prime}3^{\prime})
\\
&\times\frac{V(p^{\prime}_a-p_a) V(p^{\prime}_c-p_c)\delta_{p_a+p_b+p_c,p^{\prime}_a+p^{\prime}_b+p^{\prime}_c}}{\epsilon_b+\epsilon_c-\epsilon_{c^{\prime}}-\epsilon_{b+c-c^{\prime}}+i0}
\\
&\times\delta_{\sigma_a,\sigma_{a^{\prime}}}\delta_{\sigma_b,\sigma_{b^{\prime}}}\delta_{\sigma_c,\sigma_{c^{\prime}}},
\end{split}
\label{Eq:matrix-element-general-with spin}
\end{equation}  
where
\begin{equation}
\begin{split}
\mathcal{P}(123)&=
\\
&\{(123)^+,(231)^+,(312)^+,(132)^-,(213)^-,(321)^-\} 
\end{split}
\nonumber
\end{equation}
denotes the set of permutations and the superscript of each element denotes its sign, $\mathrm{sgn}(123)$. 
In the case of a parabolic band, the energy denominator in Eq. (\ref{Eq:matrix-element-general-with spin}) can be rewritten as follows:
\begin{equation}
\begin{split}
\epsilon_b+\epsilon_c-\epsilon_{c^{\prime}}-\epsilon_{b+c-c^{\prime}}=&-\frac{1}{m}(p_b-p^{\prime}_c)(p_c-p^{\prime}_c)
\\
&=\frac{(p_a-p^{\prime}_c)(p_b-p^{\prime}_c)(p_c-p^{\prime}_c)}{m(p^{\prime}_c-p_a)}.
\end{split}
\label{Eq:denTrans}
\end{equation}

The crucial simplification comes from the observation that the numerator in Eq. (\ref{Eq:denTrans}), when evaluated on the mass shell,  is actually invariant with respect to all the permutations of particles in the initial {\it and} final state:
\begin{equation}
(p_a-p^{\prime}_c)(p_b-p^{\prime}_c)(p_c-p^{\prime}_c)=\frac{q^3}{4}(\cos3\phi-\cos3\phi^\prime).
\label{Eq:Inv}
\end{equation}
To derive Eq. (\ref{Eq:Inv}) we have used  the parametrizations \eqref{Eq:1DParam1} and \eqref{Eq:1DParam2}. 
Equation (\ref{Eq:Inv}) leads to 
\begin{widetext}
\begin{equation}
M_{123}^{1^{\prime}2^{\prime}3^{\prime}}=\frac{4m}{q^3(\cos3\phi-\cos3\phi^\prime)}\sum_{\substack{(abc) \in \mathcal{P}(123)\\(a^{\prime}b^{\prime}c^{\prime})\in \mathcal{P}(1^{\prime}2^{\prime}3^{\prime})}}\hspace{-0.4cm}\mathrm{sgn}(123)\;\mathrm{sgn}(1^{\prime}2^{\prime}3^{\prime})\;
q_{c^\prime,a}V(q_{a^\prime, a}) V(q_{c^\prime, c})\delta_{p_a+p_b+p_c,p^{\prime}_a+p^{\prime}_b+p^{\prime}_c}
\delta_{\sigma_a,\sigma_{a^{\prime}}}\delta_{\sigma_b,\sigma_{b^{\prime}}}\delta_{\sigma_c,\sigma_{c^{\prime}}}.
\label{Eq:matrix-element-general-with-spin-Final}
\end{equation}  
\end{widetext}
Here we have introduced the shorthand notation $q_{a^\prime, b}=p_a^\prime-p_b$. 

Equation (\ref{Eq:matrix-element-general-with-spin-Final}) constitutes the basis for our discussion of the transition probability $w_q(\phi, \phi^\prime)$ in Sec. \ref{Sec:1D}. In the spinpolarized case it can be further simplified by omitting the spin indices (and adjusting the combinatorial prefactor). This leads to Eq. \eqref{Eq:MTrippleFinal} of the main text.

\section{Low-energy regime in 1D}
\label{App:1D-low-energy}

\begin{figure}
\centering
\includegraphics[width=0.95\columnwidth]{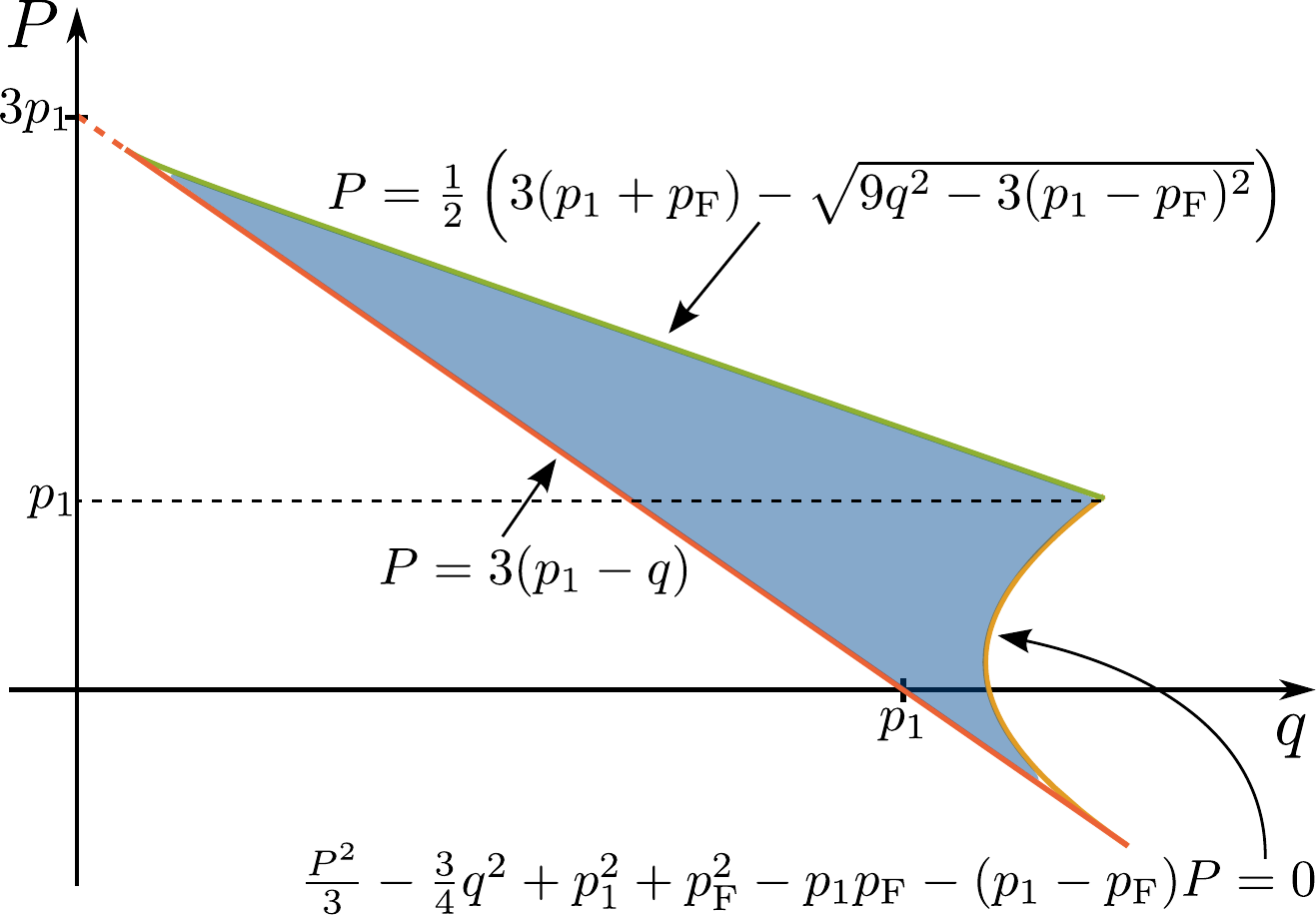}
\caption{Integration region in the $q$-$P$-plane for low-energy triple collisions in 1D, as constrained by the Fermi functions associated with incoming particles.}
\label{Fig:3particle_initial_Fermi-function_low_energy}
\end{figure}

In this appendix we present a derivation of Eqs.~\eqref{Eq:1Dspinless_low_p1}  and \eqref{Eq:1D-low-energy-spinful} for 
the low-momentum limit, $p_1-\pf \ll \pf$,  of the relaxation rate for fermions interacting via a short-range potential in one spatial dimension.  Our starting point is the expression \eqref{Eq:relaxation-rate-1D-convenient} for the relaxation rate. We begin by considering the kinematic constraints dictated by the Fermi functions $\nf(\epsilon_2)$ and $\nf(\epsilon_3)$. The area in the $q$-$P$-plane compatible with these constraints is depicted in Fig.~\ref{Fig:3particle_initial_Fermi-function_low_energy}.

\begin{figure}
\centering
\includegraphics[width=0.95\columnwidth]{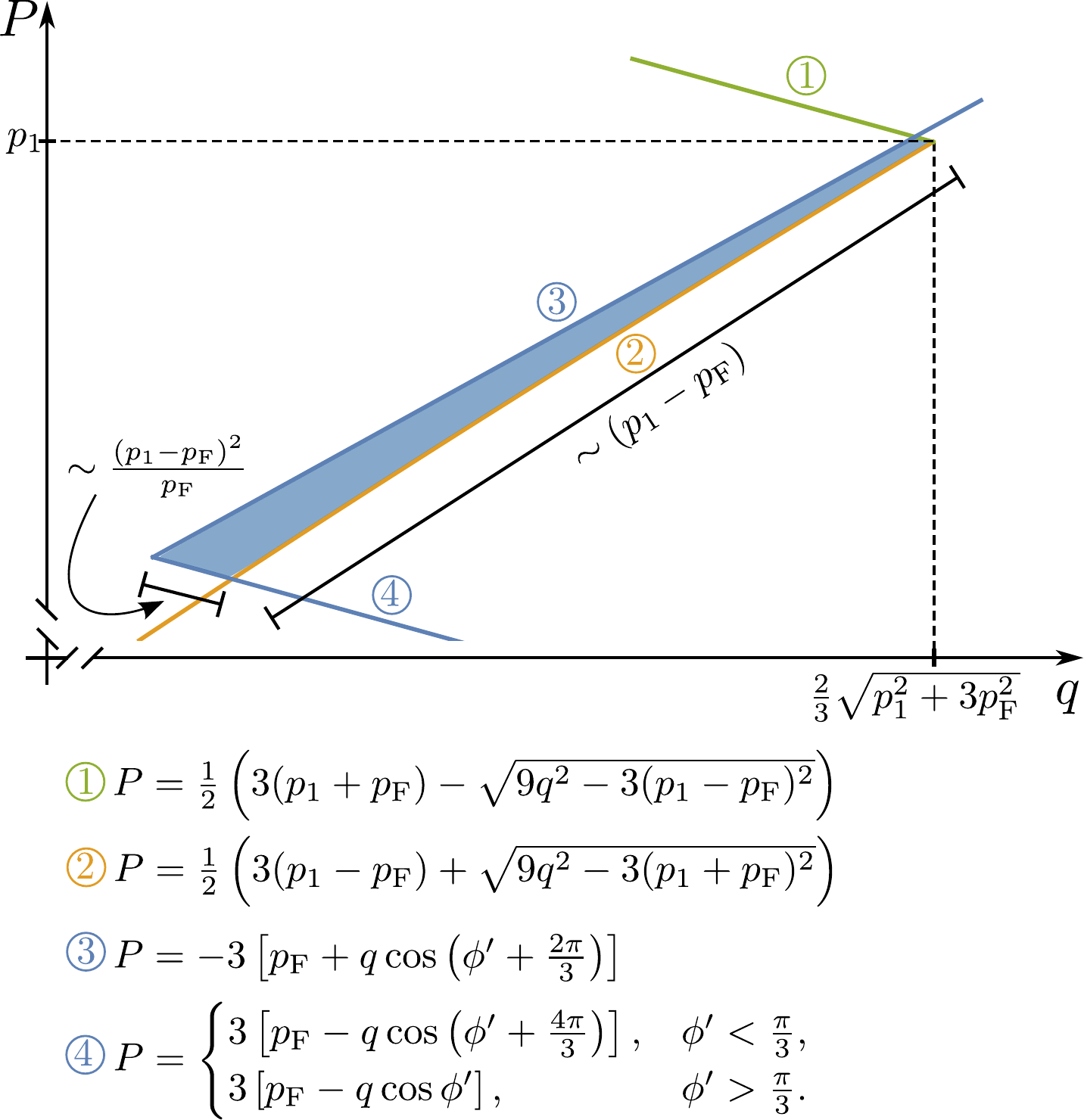}
\caption{Kinematic constraints in the $q$-$P$-plane for triple collisions in 1D at low energy. Only the upper right corner of the domain in Fig.~\ref{Fig:3particle_initial_Fermi-function_low_energy} actually contributes.\label{Fig:3particle_kinematics_low_energy}}
\end{figure}

Unlike the situation at high energies, $p_1 \gg \pf$, where energies of the particles after a scattering event ($1'$, $2'$, and $3'$)  are typically high above the Fermi sea, at low energies the Fermi functions associated with these particles further restrict the integration region in the $q$-$P$-plane. Because of the symmetry $\phi^{\prime}\to \phi^{\prime}+2\pi/3$ (interchange of particles) of the conditions resulting from the Fermi functions and of the squared matrix element, we can restrict ourselves to the interval $0<\phi^{\prime}<2\pi/3$. In the low-energy regime, $\delta=(p_1-\pf)/\pf \ll 1$, only a narrow interval 
\begin{equation}
\frac{\pi}{3}-\frac{\sqrt{3}}{4}\delta<\phi^{\prime}<\frac{\pi}{3}+\frac{\sqrt{3}}{4}\delta
\end{equation}
contributes. Furthermore, the integration domain in the $q$-$P$-plane is strongly reduced. The integration region incorporating all kinematic constraints is shown in Fig.~\ref{Fig:3particle_kinematics_low_energy}. 
As shown in the figure, only a vicinity of $q=4\pf/3$ and $P=p_1\simeq \pf$ contributes. The area of the domain in the $q$-$P$ plane scales as $(p_1-\pf)^3/\pf$. The angle $\phi$ is fixed by Eq.~\ref{Eq:phi0-1D} to $\phi_0 = \pm \pi/3 +\mathcal{O}(\delta)$. The involved particles reside close to the Fermi points, with one left mover and two right movers. 

Besides these phase-space considerations, we have to analyze the squared matrix element. As before, we assume that the momentum scale of the interaction is larger than the Fermi momentum, $q_0 \gg \pf$.  For spinless fermions we can use the form \eqref{Eq:1D-squared-matrix-element-large-q0} and find for the relaxation rate 
\begin{equation}
\begin{split}
\frac{1}{\tau_{p_1}}\sim \frac{m^3\pf^6V_0^4}{q_0^{16}}(p_1-\pf)^8, \quad p_1-\pf\ll&\pf\ll q_0,
\\
&(\mathrm{spinless})
\end{split}
\end{equation}
which is Eq.~\eqref{Eq:1Dspinless_low_p1} of the main text.

In the case of fermions with spin the behavior of the squared matrix element for $\phi\approx \pi/3$ and $\phi^{\prime}\approx \pi/3$, with a typical distance between $\phi$ and $\phi'$ of the order of $\delta=(p_1-\pf)/\pf$, is given by
\begin{equation}
w_{q}(\phi,\phi^{\prime})\sim \frac{m^2V_0^4}{q_0^4\delta^2}\sim\frac{m^2 V_0^4 \pf^2}{q_0^4(p_1-\pf)^2}.
\end{equation}
Because of the energy denominators and of the absence of the Hartree-Fock cancellation, the matrix element is strongly enhanced at low energy. Combining all the estimates of the phase space and of the matrix element, we arrive at the following result for the relaxation rate of spinful fermions at low energy in 1D:
\begin{equation}
\frac{1}{\tau_{p_1}}\sim \frac{m^3V_0^4}{q_0^4}(p_1-\pf)^2, \quad p_1-\pf \ll \pf \ll q_0 \quad (\mathrm{spinful}),
\end{equation}
 which is Eq.~\eqref{Eq:1D-low-energy-spinful} of the main text.




\begin{thebibliography}{100}

\bibitem{WeakLocalization}
G. Bergman, Phys. Rep. {\bf 107}, 1 (1984). As a more recent examples in the context of novel electronic materials see 
F. V. Tikhonenko, D. W. Horsell, R. V. Gorbachev, and A. K. Savchenko
Phys. Rev. Lett. {\bf 100}, 056802  (2008); J. G. Checkelsky, Y. S. Hor, M.-H. Liu, D.-X. Qu, R. J. Cava, and N. P. Ong Phys. Rev. Lett. {\bf 103}, 246601 (2009).

\bibitem{QHInterferometry}
Y. Ji, Y. C. Chung, D. Sprinzak, M. Heiblum, D. Mahalu,
and H. Shtrikman, Nature (London) {\bf 422}, 415 (2003);
I. Neder, M. Heiblum, Y. Levinson, D. Mahalu, and
V. Umansky, Phys. Rev. Lett. {\bf 96}, 016804 (2006).


\bibitem{EERelaxationNormal} 
See e.g. C. A. Schmuttenmaer, M. Aeschlimann, H. E. Elsayed-Ali, R. J. D. Miller, D. A. Mantell, J. Cao, and Y. Gao
Phys. Rev. B {\bf 50}, 8957(R)  (1994); 
T. Hertel, E. Knoesel, M. Wolf, and G. Ertl
Phys. Rev. Lett. {\bf 76}, 535  (1996); M. Aeschlimann, M. Bauer, S. Pawlik, W. Weber, R. Burgermeister, D. Oberli, and H. C. Siegmann
Phys. Rev. Lett. {\bf 79}, 5158  (1997).

\bibitem{EERelaxationSuperconductors} 
W. Nessler, S. Ogawa, H. Nagano, H. Petek, J. Shimoyama, Y. Nakayama, and K. Kishio,
Phys. Rev. Lett. {\bf 81}, 4480 (1998).


\bibitem{RelaxationColdAtoms} 
T. Kinoshita, T. Wenger, and D.S. Weiss, Nature (London) {\bf 440}, 900 (2006); S. Hofferberth, I. Lesanovsky, B. Fischer, T. Schumm, and J. Schmiedmayer,  Nature {\bf 449}, 324 (2007).

\bibitem{Beliaev}
S.T. Beliaev, Sov. Phys. JETP {\bf 7}, 299 (1958).

\bibitem{Tan2010}
S. Tan, M. Pustilnik, and L.I. Glazman,  Phys. Rev. Lett. {\bf 105}, 090404 (2010).

\bibitem{Ristivojevic2014}
Z. Ristivojevic and K.A. Matveev, Phys. Rev. B {\bf 89}, 180507(R) (2014).

\bibitem{Ristivojevic2016}
Z. Ristivojevic and K.A. Matveev, Phys. Rev. B {\bf 94}, 024506 (2016).

\bibitem{Relaxation1DExp}
Y.-Fu Chen, T. Dirks, G. Al-Zoubi, N. O. Birge, and N. Mason, Phys. Rev. Lett. {\bf 102}, 036804 (2009); G. Barak, H. Steinberg, L.N. Pfeiffer, K.W. West, L.  Glazman, F. von Oppen, and A. Yacoby, Nature Phys. {\bf 6}, 489 (2010). 

\bibitem{RelaxationHallExp}  
C. Altimiras, H. le Sueur, U. Gennser, A. Cavanna, D. Mailly, and F. Pierre, Nature Phys. {\bf 6}, 34 (2010); Phys. Rev. Lett. {\bf 105}, 226804 (2010);
H. le Sueur, C. Altimiras, U. Gennser, A. Cavanna, D. Mailly, and F. Pierre, Phys. Rev. Lett. {\bf 105}, 056803 (2010); N. Paradiso, S. Heun, S. Roddaro, L. Sorba, F. Beltram, and G. Biasiol, Phys. Rev. B {\bf 84}, 235318 (2011); E.V. Deviatov, A. Lorke, G. Biasiol, and L. Sorba, Phys. Rev. Lett. {\bf 106}, 256802 (2011); M.G. Prokudina, S. Ludwig, V. Pellegrini, L. Sorba, G. Biasiol, and V.S. Khrapai, Phys. Rev. Lett.  {\bf 112}, 216402 (2014). 


\bibitem{Beidenkopf2017}
J. Reiner, A. K. Nayak, N. Avraham, A. Norris, B. Yan, I. C. Fulga, J.-H. Kang, T. Karzig, H. Shtrikman, and H. Beidenkopf, Phys. Rev. X {\bf 7}, 021016 (2017).

\bibitem{Lunde07} A.M.~Lunde, K.~Flensberg, and L.I. Glazman, Phys. Rev. B
{\bf 75}, 245418 (2007).

\bibitem{Khodas2007} M. Khodas, M. Pustilnik, A. Kamenev, L.I. Glazman,  Phys.
Rev. B {\bf 76}, 155402 (2007).

\bibitem{imambekov09} A.~Imambekov and L.I.~Glazman, Science {\bf 323}, 228
(2009); Phys. Rev. Lett. {\bf 102}, 126405 (2009).

\bibitem{imambekov11} A.~Imambekov, T.L.~Schmidt, and L.I.~Glazman, Rev. Mod.
Phys {\bf 84}, 1253 (2012).

\bibitem{micklitz11} T.~Micklitz and A.~Levchenko, Phys. Rev. Lett. {\bf 106},
196402 (2011). 

\bibitem{ristivojevic13} Z.~Ristivojevic and K.A.~Matveev, Phys. Rev. B {\bf
87}, 165108 (2013). 

\bibitem{apostolov13} S.~Apostolov, D.E.~Liu, Z.~Maizelis, and
A.~Levchenko, Phys. Rev. B {\bf 88}, 045435 (2013). 

\bibitem{Lin2013} J. Lin, K. A. Matveev, M. Pustilnik,  Phys. Rev. Lett. {\bf
110}, 016401 (2013).

\bibitem{MatveevFurusaki2013} K.A. Matveev, A. Furusaki,  Phys. Rev. Lett. {\bf
111}, 256401 (2013).

\bibitem{PGM2013} I.V. Protopopov, D.B. Gutman, M. Oldenburg, and A.D. Mirlin,
Phys. Rev. B {\bf 89}, 161104(R) (2014). 

\bibitem{Matveev2013} T.~Micklitz, J.~Rech, and K.A.~Matveev, Phys. Rev. B
{\bf 81}, 115313 (2010); K.A. Matveev, J. Exp. Theor. Phys. {\bf 117}, 508
(2013).

\bibitem{Dmitriev2012} A.P. Dmitriev, I.V. Gornyi, D.G. Polyakov, Phys. Rev.
B {\bf 86}, 245402 (2012).

\bibitem{Protopopov2014}
I.V.  Protopopov, D.B. Gutman,  and A.D. Mirlin,  Phys. Rev. B {\bf 90}, 125113 (2014).

\bibitem{Gangardt2014}  M. Arzamasovs, F. Bovo, and D. M. Gangardt,  Phys. Rev. Lett. {\bf 112}, 170602 (2014).

\bibitem{protopopov13} I.V.~Protopopov, D.B.~Gutman, P.~Schmitteckert, and
A.D.~Mirlin, Phys. Rev. B {\bf 87}, 045112 (2013). 

\bibitem{KarzigGlazmanOppen10} T. Karzig, L. I. Glazman, and F. von Oppen, Phys. Rev. Lett. {\bf 105}, 226407 (2010).

\bibitem{AltshulerAronov} 
B. L. Altshuler and A. G. Aronov,  Electron-electron
interaction in disordered conductors, in {\it Electron-electron
interactions in disordered systems}, edited
by A. L. Efros and M. Pollak, (North Holland, Amsterdam
1985).

\bibitem{GMP2005} I.V. Gornyi, A.D. Mirlin, and D.G. Polyakov, Phys. Rev. Lett. {\bf 95}, 206603 (2005).

\bibitem{BAA2006}
D. M. Basko, I. L. Aleiner,  and B. L. Altshuler,  Annals of Physics {\bf 321}, 1126 (2006).

\bibitem{gornyi17} I.V. Gornyi, A.D. Mirlin, M. M\"uller, and D.G. Polyakov, Ann. Phys. (Berlin) {\bf 529}, 1600365 (2017). 

\end{thebibliography}
\end{document}